\documentclass[letterpaper,11pt]{article}
\pdfoutput = 1

\usepackage{euscript}
\usepackage{amssymb}
\usepackage{amsfonts}
\usepackage{amsbsy}
\usepackage{epsfig}
\usepackage{amsthm}
\usepackage{amscd}
\usepackage{amstext}
\usepackage{verbatim}
\usepackage{amsmath}
\usepackage{cancel}
\usepackage{capt-of}
\usepackage{empheq}

\usepackage[pdftex]{hyperref}

\hypersetup{
	colorlinks=true,
	linktoc=page,
	linkcolor=blue,
	citecolor=blue,
	urlcolor=blue
	}

\def\ben{\begin{equation}}
\def\een{\end{equation}}

\let\a=\alpha  \let\g=\gamma

\let\pa=\partial
\def\be{\begin{equation}}
\def\ee{\end{equation}}
\def\beq{\begin{equation}}
\def\eeq{\end{equation}}
\def\ba{\begin{array}}
\def\ea{\end{array}}

\def\dalemb#1#2{{\vbox{\hrule height .#2pt
       \hbox{\vrule width.#2pt height#1pt \kern#1pt
               \vrule width.#2pt}
       \hrule height.#2pt}}}

\newcommand{\bea}{\begin{eqnarray}}
\newcommand{\eea}{\end{eqnarray}}
\newcommand{\tr}{{\rm tr} }

\def\R{{{\Bbb R}}}

\newcommand{\dd}{{\rm d}}

\DeclareMathOperator{\diag}{diag}

\begin{document}

\begin{center}

{ \Large {\bf
Entropy production, viscosity bounds and bumpy black holes
}}

\vspace{1cm}

Sean A. Hartnoll$^{1}$, David M. Ramirez$^{1}$  and Jorge E. Santos$^{2}$

\vspace{1cm}

{\small
$^{1}${\it Department of Physics, Stanford University, \\
Stanford, CA 94305-4060, USA }}

\vspace{0.5cm}

{\small
$^{2}${\it Department of Applied Mathematics and Theoretical Physics, \\
University of Cambridge, Wilberforce Road, \\
Cambridge CB3 0WA, UK}}

\vspace{1.6cm}

\end{center}

\begin{abstract}

The ratio of shear viscosity to entropy density, $\eta/s$, is computed in various holographic geometries that break translation invariance (but are isotropic). The shear viscosity does not have a hydrodynamic interpretation in such backgrounds, but does quantify the rate of entropy production due to a strain. Fluctuations of the metric components $\delta g_{xy}$ are massive about these backgrounds, leading to $\eta/s < 1/(4\pi)$ at all finite temperatures (even in Einstein gravity). As the temperature is taken to zero, different behaviors are possible. If translation symmetry breaking is irrelevant in the far IR, then $\eta/s$ tends to a constant at $T=0$. This constant can be parametrically small. If the translation symmetry is broken in the far IR (which nonetheless develops emergent scale invariance), then $\eta/s \sim T^{2 \nu}$ as $T \to 0$, with $\nu \leq 1$ in all cases we have considered. While these results violate simple bounds on $\eta/s$, we note that they are consistent with a possible bound on the rate of entropy production due to strain.

\end{abstract}

\thispagestyle{empty}

\pagebreak
\setcounter{page}{1}

\newpage

\section{Introduction}

\subsection{Motivation}

The proposal of a fundamental bound on the ratio of shear viscosity to entropy density has stimulated much work over the past decade. The bound was initially formulated as \cite{Kovtun:2004de}
\be\label{eq:KSS}
\frac{\eta}{s} \geq \frac{1}{4 \pi} \frac{\hbar}{k_B} \,.
\ee
Three arguments were given for such a bound. Firstly, an energy-time uncertainty principle argument, applied to the scattering rate in a Boltzmann equation (weak coupling) description, suggested such a bound \cite{Kovtun:2004de}. Secondly, strong coupling results from the simplest holographic theories \cite{Policastro:2001yc, Kovtun:2003wp} `universally' saturated the bound (\ref{eq:KSS}). Thirdly, the bound was in agreement with known experimental data \cite{Kovtun:2004de}. This last fact has become more intriguing with subsequent experimental results for the viscosity from the quark-gluon plasma as well as in cold Fermi gases at unitarity being consistent with the existence of a bound. See for instance \cite{Adams:2012th}.

A potential bound along the lines of (\ref{eq:KSS}) is exciting for at least two reasons. Firstly, it suggests the possibility that basic quantum mechanical (and thermodynamic) principles may control the behavior of physically important quantities, beyond any weak coupling `quasiparticle' description. For instance, some of the arguments leading to the viscosity bound (\ref{eq:KSS}) can be adapted to suggest an understanding of `bad metals' in terms of bounds on the electrical conductivity \cite{Hartnoll:2014lpa, Lucas:2015lna, Grozdanov:2015qia}. Secondly, because the bound might be saturated by the holographic dynamics of black hole horizons, it points towards an understanding of black holes as the `most extreme' quantum systems in nature. Such a formulation heuristically connects with other extreme aspects of black holes, such as fast scrambling \cite{Sekino:2008he, Maldacena:2015waa}.

Controlled theoretical counterexamples to the originally proposed bound (\ref{eq:KSS}) have been found in theories with higher derivative gravity duals \cite{Kats:2007mq, Buchel:2008vz}. In certain special cases, however, it can be shown that the viscosity to entropy ratio remains bounded from below \cite{Brigante:2008gz, Hofman:2008ar}. See \cite{Cremonini:2011iq} for an overview of these and related results. Within this set of ideas, it remains plausible that a bound along the lines of (\ref{eq:KSS}) might hold -- with a different numerical factor --  at least for some large class of consistent quantum theories.

A more dramatic violation of (\ref{eq:KSS}) has been found in anisotropic systems \cite{Rebhan:2011vd, Mamo:2012sy, Jain:2014vka, Critelli:2014kra, Jain:2015txa}. In an anisotropic system there are multiple components of the shear viscosity, that can behave differently. In the models of \cite{Rebhan:2011vd, Mamo:2012sy, Jain:2014vka, Critelli:2014kra, Jain:2015txa} it was found that, as $T \to 0$ with an anisotropy scale held fixed, one component of the shear viscosity behaved as
\be\label{eq:anisotropic}
\frac{\eta_\perp}{s} \sim T^\alpha \,,
\ee
with the exponent $0 < \alpha \leq 2$. This gives a parametric violation of the bound (\ref{eq:KSS}) at low temperatures. That said, it is not obvious a priori that $\eta_\perp/s$ is the `right' quantity to bound in anisotropic systems.\footnote{On the experimental front, a very low spin diffusivity has been measured in an anisotropic cold Fermi gas \cite{anis}. This may be related to the discussion here because, in a translation invariant system, the shear viscosity over entropy density is essentially the momentum diffusivity \cite{Son:2007vk}
.} The computations in \cite{Rebhan:2011vd, Mamo:2012sy, Jain:2014vka, Critelli:2014kra, Jain:2015txa} show that $\eta_\perp$ is `universally' given by the geometry of the horizon, in the same sense that leads to the `universal' saturation of (\ref{eq:KSS}) by isotropic horizons \cite{Iqbal:2008by}. Thus the holographic anisotropic geometries could be argued to saturate a different bound, in which the entropy density is rescaled by the degree of anisotropy in the low energy physics. Such considerations beg the question: what exactly are these bounds trying to say? What are the rules of the game? Perhaps the bounds have not yet been formulated in the most transparent way. In this paper we hope to shed some light on these questions.

\subsection{Summary}

The main content of this paper involves new holographic results for $\eta/s$. We will furthermore discuss a possible
interpretation of these and previous results in terms of bounds.

In section \ref{sec:formula} we show that $\eta/s$ is equal to the increase of the logarithm of the entropy density per `Planckian time' \cite{zaanen}, when the system is perturbed by a linear in time source for the background metric $\delta g_{xy}^{(0)} = t \, T$ (i.e. a strain). One advantage of this description of $\eta/s$ -- that will be important for our purposes -- is that it is independent of any role that $\eta$ may play in hydrodynamics. It remains true in the presence of translation symmetry breaking, where momentum is not conserved and $\eta$ is not a hydrodynamic quantity. A potentially more significant advantage of the reformulation is the following. In certain systems a (typically momentum) scale $\Delta$ survives in the zero temperature IR fixed point description. One might then consider bounding the entropy production per Planckian time in the presence of a different, temperature-independent, source $\delta g_{xy}^{(0)} = t \, \Delta$. This will lead to a different, weaker and temperature dependent, bound on $\eta/s$ that is compatible with the anisotropic results (\ref{eq:anisotropic}), as well as for the new results for $\eta/s$ that we find in this paper.

In the remaining sections we will compute $\eta/s$ in isotropic but non-translation invariant holographic spacetimes. We consider certain simple backgrounds that can be studied without the use of PDEs. We will see that $\eta/s$ can become arbitrarily small in these theories as the temperature is lowered. In some cases, the behavior is rather similar to the anisotropic result (\ref{eq:anisotropic}). However, unlike in that case, $\eta/s$ will not be a `universal' horizon quantity. Instead it must be extracted from properties of a non-normalizable perturbation of the entire spacetime. Specifically, we will find
\begin{itemize}
\item For neutral and charged `linear axion' models, $\eta/s \sim T^{2\nu}$ as $T \to 0$, with $0 \leq \nu \leq 1$. Translation invariance is broken at the $T=0$ IR fixed point in these models. The zero temperature IR geometry is $AdS_2 \times \R^2$.

\item Neutral `Q-lattice' models are found to have $\eta/s \to \text{const.}$ as $T \to 0$. For lattices with large amplitude, the $\text{const.} \ll 1$. Translation invariance is restored at the $T=0$ IR fixed point of these models. The zero temperature IR geometry is $AdS_4$.

\item Charged Q-lattices can show emergent translation invariance in the $T=0$ IR fixed point or else exhibit strongly non-translation invariant (`insulating') low energy physics. The former case has $\eta/s \to \text{const.}$ as $T \to 0$, and the zero temperature IR geometry is $AdS_2 \times \R^2$. The insulating cases are not understood analytically, but appear to show a slow (possibly power law) decay of $\eta/s$ to zero as $T \to 0$.

\end{itemize}

The physical picture behind these results is as follows. Without translation symmetry in the IR, there is no privileged dynamical mode, and $\eta/s$ simply depends on the number of low energy quantum critical excitations that overlap with the $T^{xy}$ operator. This leads to the power law $\eta/s \sim T^{2 \nu}$. With translation invariance emerging in the IR, a collective diffusive mode leads to a constant $\eta/s$ as $T \to 0$. However, with translation symmetry strongly broken at intermediate energy scales, this mode can have very small overlap with the microscopic (UV) $T^{xy}$ operator, leading to $\eta/s \ll 1$.

\section{Shear viscosity and entropy production}
\label{sec:entropy}

Define the shear viscosity in a relativistic QFT as
\be\label{eq:eta}
\eta \equiv \lim_{\omega \to 0} \frac{1}{\omega} \, \text{Im} \, G^R_{T^{xy} T^{xy}}(\omega,k=0) \,.
\ee
Here $T^{xy}$ are spatial parts of the energy-momentum tensor. In the absence of sources that break translation invariance, then $T^{xy}$ is the current corresponding to conservation of momentum $T^{ty}$. In particular, in that case, equation (\ref{eq:eta}) is a Kubo formula for the `transverse momentum conductivity'. In hydrodynamics, $\eta$ then gives the momentum diffusivity via the Einstein relation $D = \eta/(sT)$ \cite{Son:2007vk}.

In a relativistic QFT, however, the operator $T^{xy}$ is well-defined in any background, including those that break translation invariance. For instance, the theory can be placed on an inhomogeneous background metric. It follows that, even in the absence of hydrodynamics, $\eta$ retains its essential interpretation: it quantifies the rate of entropy production when the equilibrium state is subjected to a slowly varying homogeneous source (strain) $\delta g_{xy}^{(0)}$. Specifically, take the source to be linear in time
\be\label{eq:growth}
\delta g_{xy}^{(0)} = t \, c^{(0)} \,.
\ee
Here $c^{(0)}$ is a constant. The rate of entropy density production is then
\be\label{eq:sdot}
\dot s = \frac{\eta}{T} \left(c^{(0)}\right)^2 \,.
\ee
Here $\eta$ is defined as in (\ref{eq:eta}). The above formula follows from simple manipulations with linear in time sources, see e.g. \cite{Donos:2014cya},\footnote{There is an additional contact term contribution to the entropy production, proportional to the susceptibility $G^R_{T^{xy} T^{xy}}(0,0)$, which is nonzero in the cases we consider below. This is an artifact of the time dependent source growing to the far past and future. If, instead, a sufficiently smooth regularized source is used, then the contact term vanishes upon time averaging. The rates of entropy production we discuss should strictly be considered as time-averaged rates.} combined with the standard formula for the rate $\dot w$ at which external fields do work on a system (see e.g. \cite{Hartnoll:2009sz}), with $\dot w = T \dot s$.

The expression (\ref{eq:sdot}) is suggestively re-written as\footnote{The natural appearance of the logarithm of the entropy was noted first by Brian Swingle. Swingle further noted that the logarithm of the entropy possibly suggests a connection to the physics of scrambling.}
\be\label{eq:logs}
\frac{1}{T} \frac{d \log s}{dt} = \frac{\eta}{s} \left(\frac{c^{(0)}}{T}\right)^2 \,.
\ee
The ratio $c^{(0)}/T$ is dimensionless and, in a first pass, is naturally taken to be unity (i.e. choose $c^{(0)} = T$). In that case, we see that the ratio of shear viscosity to entropy density \cite{Kovtun:2004de} gives the increase of the logarithm of the entropy density over a `Planckian time' \cite{zaanen}
\be\label{eq:pl}
t_\text{Pl} \equiv \frac{\hbar}{k_B T} \,.
\ee
Here we have momentarily restored factors of the Planck and Boltzmann constants.
The important point for our purposes is that the relation (\ref{eq:logs}) remains true in the absence of translation invariance, so long as the operator $T^{xy}$ is defined.

The above paragraph shows that the viscosity bound (\ref{eq:KSS}) is equivalent to a bound on the entropy production per Planckian time due to a specific source $\delta g_{xy}^{(0)} = t \, T$. Namely
\be\label{eq:entropy}
t_\text{Pl} \; \frac{d \log s}{dt} \gtrsim 1\,.
\ee
We will not be attempting to fix numerical prefactors in this work, although we will characterize circumstances in which numerical prefactors in the viscosity can become very small.
In addition to excising the bound from its original hydrodynamical context, this formulation of a viscosity to entropy density bound suggests a generalization that we now describe.

In a theory with a full emergent IR scaling symmetry at zero temperature (in holography this could be, for example, an $AdS_{d+1}$ Poincar\'e horizon or it might be an extremal $AdS_2 \times \R^{d-1}$ horizon) the most relevant scale is indeed the temperature, and therefore the choice of source $\delta g_{xy}^{(0)} = t \, T$ is natural. However, in other circumstances, an additional scale $\Delta$ can survive into the far IR, even if the theory is gapless. This is best thought of as a momentum rather than energy scale (the latter, by definition, have decoupled in the low energy IR fixed point theory). Interesting examples where this can happen are cases where translation invariance is broken in the far IR. In holographic models this has recently been realized in three different ways: (i) disordered fixed points in which scaling only emerges in spatially averaged quantities \cite{Hartnoll:2014cua, Hartnoll:2015faa, Hartnoll:2015rza}, (ii) solutions with $z=\infty$ scaling in which space does not scale and hence spatial inhomogeneity is compatible with scaling \cite{Hartnoll:2014gaa} and (iii) cases in which the bulk matter background breaks scaling but the energy momentum tensor, and hence the metric, are scale invariant, e.g. \cite{Donos:2012js, Donos:2013eha, Andrade:2013gsa}. While the third case has some artificial features, it is much simpler to work with and therefore we will use variants of those models in this first study.

When a scale $\Delta$ is present in the zero temperature IR theory, at the lowest temperatures one could consider bounding the entropy production in the presence of a different source: $\delta g_{xy}^{(0)} = t \, \Delta$. With this source, the entropy production per Planckian time is given by
\be\label{eq:abab}
\frac{1}{T} \frac{d \log s}{dt} = \frac{\eta}{s} \left(\frac{\Delta}{T}\right)^2 \,.
\ee
It is interesting to impose the bound on entropy production (\ref{eq:entropy}), now viewed as a principle in its own right, with this new source. From (\ref{eq:abab}) one obtains the new viscosity bound
\be\label{eq:newbound}
\frac{\eta}{s} \gtrsim \left(\frac{T}{\Delta}\right)^2 \qquad \text{as} \qquad \frac{T}{\Delta} \to 0 \,.
\ee
This bound is weaker than bounds of the type (\ref{eq:KSS}), and therefore does not contradict them. It has the virtue of being satisfied by all of the anisotropic results quoted in (\ref{eq:anisotropic}) above, and it will also be satisfied by the cases we describe below.\footnote{The anisotropic geometries considered in \cite{Rebhan:2011vd, Mamo:2012sy, Jain:2014vka, Critelli:2014kra, Jain:2015txa} have matter fields that break translation invariance; they are less symmetric instances of the kind of spacetimes we will be considering shortly.} In fact, in several cases the temperature scaling will saturate (\ref{eq:newbound}). The weaker bound (\ref{eq:newbound}) should be relevant when a scale $\Delta$ survives in the IR fixed point theory.

The idea that has been outlined above can be summarized as follows: The large amount of previous work on the shear viscosity in holographic theories that found $\eta/s \sim 1$ and the work in \cite{Rebhan:2011vd, Mamo:2012sy, Jain:2014vka, Critelli:2014kra, Jain:2015txa} and in the present paper that finds $\eta/s \sim T^{2\nu}$, with $0 \leq \nu \leq 1$, at low temperatures (as well as other behaviors) in certain models, can be reconciled from the standpoint of a bound (\ref{eq:entropy}) on the rate of entropy production. The entropy is produced due to a strain acting on the thermal equilibrium state. The suggestion is that when a (typically momentum) scale is present in the low energy theory, the correct bound on entropy production at low temperatures is in the presence of the source $\delta g_{xy}^{(0)} = t \, \Delta$. A more satisfactory description should ultimately combine the different bounds into one, perhaps by bounding entropy production with a source along the lines of $\delta g_{xy}^{(0)} = t \, \sqrt{\Delta^2 + T^2}$, cf. \cite{analytis}, or indeed $\delta g_{xy}^{(0)} = t  \, \Delta(T)$. This will not be attempted here.

We do not have an argument for why the particular quantity appearing on the left hand side of (\ref{eq:entropy}) should be bounded. We also do not have a precise definition of the scale $\Delta$ that should be used in a given model. The comments above are therefore rather tentative. Nonetheless, we have found these observations suggestive and a useful framework for interpreting the concrete results that follow in the remainder of the paper. In particular, we have not found any obvious counterexamples to (\ref{eq:newbound}).

\section{A (weaker) horizon formula for \texorpdfstring{$\eta/s$}{test}}
\label{sec:formula}

In a translation invariant background, the ratio of the shear viscosity to entropy density can be evaluated directly from horizon data alone \cite{Kovtun:2003wp, Iqbal:2008by}. This is unlikely to remain true once translation invariance is broken, as the shear viscosity in this case is no longer associated with a conservation law. Methods such as those in \cite{Donos:2015gia} will most likely not apply. However, a weaker `horizon formula' for $\eta/s$ exists in certain cases as we now discuss.

The objective is to evaluate the correlator in (\ref{eq:eta}). Correlators of $T^{xy}$ are obtained from perturbations $\delta g_{xy}$ of the metric in the bulk, according to the usual holographic dictionary \cite{Hartnoll:2009sz}. In particular, we will consider backgrounds where the metric and energy-momentum tensor take the form
 \bea
    & \displaystyle \dd s^2 = {-} g_{tt}(r) \dd t^2 +
      g_{rr}(r) \dd r^2 + g_{xx}(r) \dd x^i \dd x^i \, , & \nonumber \\
    &  \displaystyle T_{\mu\nu} = \diag \Big(T_{tt}(r), T_{rr}(r), T_{xx}(r), \dotsc, T_{xx}(r)\Big)\, . & \label{eq:back}
  \eea
In particular, the metric and energy-momentum tensor are homogeneous and isotropic in the field theory directions. Crucially, however, we will not assume that the matter fields sourcing the energy-momentum tensor are homogeneous.

Taking the perturbation
\be
\left(\delta g\right)^x{}_y = h(r) e^{- i \omega t} \,,
\ee
about the background (\ref{eq:back}) leads to a wave equation with radially varying mass
  \begin{align}
    0 ={}& \frac{1}{\sqrt{{-} g}} \partial_r \left( \sqrt{{-}
        g} g^{rr} \partial_r \, h \right) +
    \left[g^{tt} \omega^2 - m(r)^2 \right]h\,
    . \label{eq:hxyeq}
  \end{align}
This mode decouples from other perturbations for the background above. The mass squared is (we will be taking $16 \pi G_N$ to multiply the entire action, and therefore it does not appear in the equations of motion)
  \begin{align}
    m(r)^2 ={}& g^{xx} T_{xx} - \frac{\delta
        T_{xy}}{\delta g_{xy}} \, . \label{eq:mass}
  \end{align}
  It is easy to check that this mass vanishes for Schwarzschild and
  Reissner-Nordstr\"om backgrounds. When the mass vanishes, then
  the arguments of \cite{Kovtun:2003wp, Iqbal:2008by} apply. That is,
  the ratio of shear viscosity to entropy density is given by horizon data and takes the `universal' value $\eta/s = 1/(4\pi)$ \cite{Kovtun:2004de, Son:2007vk}. We will, however, be interested in cases in which the mass does not vanish. The precise form of $m(r)$ is not important for the general results of this section.
  
An elegant general formula for correlators of the form (\ref{eq:eta}) has been given in \cite{Lucas:2015vna}. Holographic formulae of the sort we are about to quote have appeared for some time, see e.g. \cite{Chakrabarti:2010xy, Jain:2015txa}, but the derivation in \cite{Lucas:2015vna} -- very recently reviewed in \cite{Davison:2015taa} -- is especially crisp and general. We shall simply quote the result. Let $h_o(r)$ be the solution to the wave equation (\ref{eq:hxyeq}) at zero frequency (i.e. with $\omega = 0$) that (i) is regular on the horizon $r=r_+$ and (ii) goes like $h_o = 1$ near the boundary as $r \to \infty$. The latter condition simply means that we fix the coefficient of the non-normalizable mode near the boundary to one (in coordinates where the metric components $g_{tt}$ and $g_{xx}$ in (\ref{eq:back}) go like $r^{2}$ near the boundary). In terms of the solution $h_o(r)$ one finds
  \begin{align}
\lim_{\omega \to 0} \frac{1}{\omega} \, \text{Im} \, G^R_{T^{xy} T^{xy}}(\omega,k=0)  ={}&
    \frac{\sqrt{\gamma(r_+)}}{16 \pi G_N} h_o(r_+)^2
    = \frac{s}{4\pi} h_o(r_+)^2\, .
  \end{align}
  Here $\gamma(r_+)$ is the determinant of the spatial metric on the horizon.
  For the last equality we have noticed the factor of the entropy density $s =
  \sqrt{\gamma(r_+)}/4G_N$. From the definition of the shear viscosity
  (\ref{eq:eta}) we therefore have
  \begin{align}
    \frac{\eta}{s} ={}& \frac{1}{4\pi} h_o(r_+)^2\, . \label{eq:h2}
  \end{align}
  If the mass vanishes, then the solution to (\ref{eq:hxyeq}) at zero frequency is simply $h_o = 1$ everywhere \cite{Iqbal:2008by}. The `universal' result $\eta/s = 1/(4\pi)$ is then obtained from (\ref{eq:h2}). More generally, with nonvanishing mass, the `horizon formula' (\ref{eq:h2}) is less powerful. In particular one must solve a differential equation everywhere in the bulk. However, it still offers a considerable simplification: the equation can be solved directly with $\omega = 0$.
  
From expression (\ref{eq:h2}) for the viscosity to entropy density ratio and the differential equation (\ref{eq:hxyeq}) we can conclude that if $m^2 > 0$ everywhere, then $\eta/s$ is necessarily lower than $1/(4\pi)$. The argument is as follows. Assume throughout that $m^2>0$. Then, assume that $h_o(r_+)>0$, so that $h_o$ is positive on the horizon. In the differential equation (\ref{eq:hxyeq}), $\sqrt{-g}$ is regular on the horizon whereas $g^{rr}$ vanishes and increases away from the horizon. We are taking the radial coordinate to be in the range $r_+ < r < \infty$. It follows that $h_o'(r_+)>0$ and thus $h_o$ increases away from the horizon. Suppose that $h_o$ were to stop increasing at some point (and assume $h_o$ is continuous). This would would require $\pa_r h_o$ to vanish at that point, at which $h_o$ will still be positive. However, the full equation implies that if $\pa_r h_o = 0$ with $h_o>0$, then $\pa_r^2 h_o > 0$. That would require the stationary point to be a minimum, which is not consistent with the fact that $h_0$ is supposed to stop increasing at that point. Therefore $h_o$ must increase monotonically from the horizon to the boundary. Therefore, if $h_o$ is unity at the boundary, $h_o(r_+) < 1$ and hence $\eta/s < 1/(4 \pi)$.

We do not have a general argument for when $m(r)^2$ in (\ref{eq:mass}) should be positive. It will turn out to be positive in all the cases we consider below. In `massive gravity', which we return to briefly in the discussion section, negative $m^2$ leads to instability \cite{Vegh:2013sk, Blake:2013owa}.

Another general argument can be made for extremal horizons. It can be explicitly verified that the formula (\ref{eq:h2}) holds without change for finite size extremal horizons. At such an extremal horizon, in a convenient gauge, $g^{rr}$ has a double zero. Suppose that $m(r_+) \neq  0$, i.e. the mass does not vanish on the horizon. Then, by solving the differential equation (\ref{eq:hxyeq}) to leading order near the horizon, one finds that the regular solution vanishes at the horizon. It follows that in these cases $\eta/s = 0$ at $T = 0$. Extremal Reissner-Nordstr\"om evades this conclusion because, as we have noted, $m=0$ everywhere in that case (the two terms in (\ref{eq:mass}) cancel). Thus extremal Reissner-Nordstr\"om has $\eta/s = 1/(4 \pi)$, see e.g. \cite{Edalati:2009bi}. However, we will see that more general extremal horizons -- in particular with broken translational invariance -- can lead to vanishing $\eta/s$ at zero temperature.

We will furthermore see below that if the mass vanishes precisely at an extremal horizon, but does not vanish everywhere in the spacetime, then $\eta/s$ will again be a constant at zero temperature, but typically not equal to $1/(4 \pi)$.

\section{Isotropic Linear Axion Models}

\subsection{The black brane solution}
\label{sec:branes}

A simple holographic framework for the study of transport with momentum relaxation was considered in \cite{Andrade:2013gsa}. These can be called linear axion models. We will focus on four-dimensional bulk spacetimes. Higher dimensional generalizations are easily constructed.

The model contains two massless scalar fields, minimally coupled to gravity according to the action
\begin{equation}
S = \frac{1}{16\pi G_N}\int \mathrm{d}^4 x\sqrt{-g}\left[R+\frac{6}{L^2}-\frac{1}{2} \left(\nabla_a \vec{\phi}\right)\cdot \left(\nabla^a \vec{\phi}\right)\right] \,.
\label{eq:action}
\end{equation}
Here $\vec{\phi}$ is a two dimensional real vector and $L$ is the AdS length scale, which henceforth we will set to one. The equations of motion derived from (\ref{eq:action}) read
\begin{subequations}
\begin{equation}
R_{ab}+3\,g_{ab} = \frac{1}{2} \left(\nabla_a \vec{\phi} \right)\cdot \left(\nabla_b \vec{\phi} \right)\,,
\end{equation}
\begin{equation}
\Box \vec{\phi}=0\,.
\end{equation}
\end{subequations}

We will focus on a simple class of solutions with an isotropic metric. Translational symmetry is broken in the matter sector of the theory. The solutions are
\begin{subequations}
\begin{equation}
\mathrm{d} s^2 = -f(r)\mathrm{d}t^2+\frac{\mathrm{d}r^2}{f(r)}+r^2 \left(\mathrm{d}x_1^2+\mathrm{d}x_2^2 \right)\,,
\label{eq:line}
\end{equation}
\begin{equation}
\phi^i = \alpha\,x_i\,,\quad\text{for}\quad i\in\{1,2\}\,,
\end{equation}
\end{subequations}
with
\begin{equation}
f(r) = r^2-\frac{\alpha^2}{2}-\frac{r_+}{r}\left(r_+^2-\frac{\alpha^2}{2}\right)\,.
\end{equation}
The conformal boundary is located at $r\to+\infty$, with the conformal boundary metric, $\mathrm{d}s^2_\partial$, being conformal to three-dimensional Minkowski spacetime
\begin{equation}
\mathrm{d}s^2_\partial = -\mathrm{d}t^2+\mathrm{d}x_1^2+\mathrm{d}x_2^2\,.
\end{equation}

There is a non-degenerate horizon located at $r=r_+>0$, so long as $|\alpha|<\sqrt{6}\,r_+$. Its associated Hawking temperature \cite{Hawking:1974rv} is
\begin{equation}
T = \frac{|f'(r_+)|}{4\pi} = \frac{1}{8\pi r_+}\left(6\,r_+^2-\alpha^2\right)\,.
\label{eq:temp}
\end{equation}
The horizon degenerates at $|\alpha| = \sqrt{6}\,r_+$, remaining regular. The energy density $\varepsilon$ of the solution is computed using standard holographic renormalisation techniques \cite{deHaro:2000vlm,Andrade:2013gsa} to be
\begin{equation}
\varepsilon = \frac{r_+}{16\pi G_N}\left(2r_+^2-\alpha^2\right)\,.
\label{eq:energy}
\end{equation}
The entropy density $s$ can be readily obtained from (\ref{eq:line}), $s = r_+^2/(4 G_N)$. We note that at extremality the energy density is negative, $\varepsilon = -|\alpha|^3 /(24 \sqrt{6} \pi  G_N)$, reminiscent of the behavior of the AdS soliton \cite{Horowitz:1998ha}.

\subsection{Results for \texorpdfstring{$\eta/s$}{test}}

Following the general discussion in section \ref{sec:formula}, the background is perturbed by the time independent mode
\begin{equation}
\delta g_{x_1x_2} = g_{x_1 x_1} h_o(r) = r^2 h_o(r)\,.
\end{equation}
This mode decouples from the remaining spacetime perturbations, and yields a simple equation for $h_o(r)$
\begin{equation}
\frac{1}{r^2}\frac{\mathrm{d}}{\mathrm{d}r}\left[r^2 f(r)\frac{\mathrm{d}}{\mathrm{d}r}h_o\right] - \frac{\alpha^2}{r^2}h_o=0\,. \label{eq:linear}
\end{equation}
In the language of section \ref{sec:formula}, this simple model has mass $m(r)^2=\alpha^2/r^2$ in equation (\ref{eq:hxyeq}). The general argument of section \ref{sec:formula} already tells us that the positivity of $m(r)^2$ implies that we are going to find $\eta/s < 1/(4 \pi)$ at all temperatures. Furthermore, the second general argument argument from section \ref{sec:formula} tells us that because $m(r_+)$ is nonzero even at extremality ($|\alpha| = \sqrt{6}\,r_+$), we know that we will have $\eta/s \to 0$ as $T \to 0$. Thus, we know a fair amount before solving any equations!

This model describes a CFT perturbed by two scales, the temperature $T$ and slope of the axion source $\alpha$. The viscosity to entropy density ratio can therefore only be a function of $T/\alpha$.
In Appendix \ref{sec:linear} we solve equation (\ref{eq:linear}) analytically in high and low temperature expansions, and at a special `self-dual' temperature. We solve the equation numerically at all temperatures. Here we summarize the results.

\begin{itemize}

\item In a high temperature expansion:
\bea
\lefteqn{4\pi\, \frac{\eta}{s} =  1+\frac{\sqrt{3}}{16 \pi }\left(1-\frac{3 \sqrt{3} \log 3}{\pi }\right)\left(\frac{\alpha }{T}\right)^2+} \label{eq:highTlinear}\\
& & \frac{9 \sqrt{3}}{512 \pi ^3}\left[1-\frac{3 \sqrt{3} \pi }{2}-5 \log 3-\frac{\sqrt{3} \log 3}{\pi }+\frac{9 \sqrt{3} \log ^2 3}{2 \pi }+\frac{2 \sqrt{3}}{\pi} \psi^{(1)}\left(\frac{1}{3}\right)\right]\left(\frac{\alpha }{T}\right)^4 +\mathcal{O}\left[\left(\frac{\alpha }{T}\right)^6\right] \,. \nonumber
\eea
Here $\psi^{(1)}$ is the derivative of the digamma function.

\item In a low temperature expansion:
\begin{equation}
4\pi \, \frac{\eta}{s} = \frac{512\pi^2}{2187\left(\frac{4}{3}-\log3\right)^2}\left(\frac{T}{\alpha}\right)^2+\mathcal{O}\left[\left(\frac{T}{\alpha}\right)^4\right]\,. \label{eq:lowTlinear}
\end{equation}
Note that the term proportional to $(T/\alpha)^3$ is zero.

\item Result at $T = \alpha/(\sqrt{8} \pi)$:
\begin{equation}
4\pi\, \frac{\eta}{s}=\frac{4}{\pi} \left| \Gamma \left(\frac{5}{4}-\frac{i \sqrt{7}}{4}\right)\right| ^4 \approx 0.325\,. \label{eq:selfduallinear}
\end{equation}
This result was found previously in \cite{Davison:2014lua}.

\item Results for general temperature are shown in Fig.~\ref{fig:scalars} in a log-log plot. The vertical axis is $4\pi \eta /s$, and the horizontal axis is $\alpha/T$. The exact and perturbative results just discussed are also shown: the red dashed line is the perturbative expansion for small $\alpha/T$, the red diamond is the result for $\alpha/T = \sqrt{8}\pi$, the dotted blue line is the perturbative result at large $\alpha/T$. The numerical data is represented by the green disks. The agreement between the analytic and numerical results is reassuring.

\begin{figure}[h]
\centering
\includegraphics[height = 0.35\textheight]{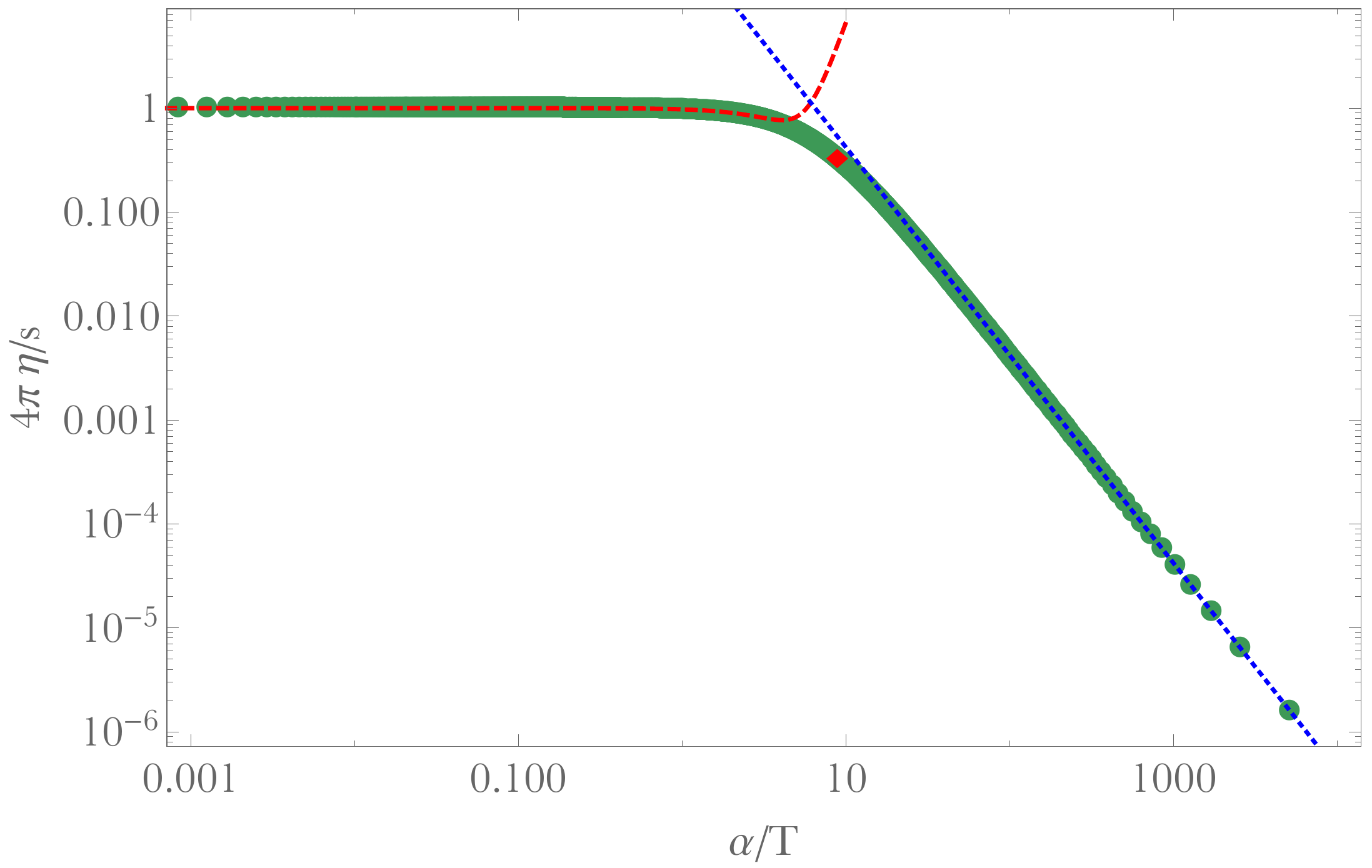}
\caption{\label{fig:scalars}  {\bf Log-log plot of $4\pi \eta /s$ as a function of $\alpha/T$}, for neutral linear axion backgrounds. The red dashed line is the perturbative expansion (\ref{eq:highTlinear}) for small $\alpha/T$, the red diamond is the exact result (\ref{eq:selfduallinear}) for $\alpha/T = \sqrt{8}\pi$, the dotted blue is the perturbative result (\ref{eq:highTlinear}) at large $\alpha/T$. The green disks are data obtained from numerically solving (\ref{eq:scalarsh}).}
\end{figure}

\end{itemize}

The results we have just obtained fit precisely into the picture outlined in section \ref{sec:entropy}. At high temperatures, the effect of translational symmetry breaking is negligible, and thus $\eta/s = 1/(4 \pi)$ pertains. At the lowest temperatures, the translation symmetry breaking scale $\Delta = \alpha$ dominates and hence we find $\eta/s \sim T^2$ as $T \to 0$.

\subsection{Nonzero charge density}
\label{sec:scale}

A simple generalization of the above setup is to allow a nonzero charge density.
This is done by adding a Maxwell field to the action (\ref{eq:action}).
Nonzero charge density solutions are characterized by an additional scale,
the chemical potential $\mu$. In particular, at zero temperature, one may tune $\alpha/\mu$. Doing so will lead us to an instructive physical picture of the low temperature scaling of the viscosity to entropy density ratio.

Because the form of the charged background and perturbation equation very much parallels those described in previous sections, we give a few details in Appendix \ref{sec:linear} and jump straight to the results here. Since the mass squared $m(r)^2$ of $(\delta g)^x{}_y$ is again equal to $\a^2/r^2 > 0$, the general argument of section \ref{sec:formula} says that we must have $\eta/s < 1/(4\pi)$ for all $\alpha > 0$. Furthermore, since on the horizon $m(r_+)^2 = \a^2/r_+^2$ is nonzero, the arguments of section \ref{sec:formula} furthermore imply that for all $\alpha >0$, $h_o$ must vanish at the horizon at $T=0$, and hence $\eta/s = 0$ at zero temperature. We can do better however, and obtain analytically the power of $T$ with which $\eta/s$ vanishes.

At $T=0$ the near horizon geometry of the spacetime is $AdS_2 \times \R^2$. The behavior of the regular perturbation $h_o$ in the near horizon region is
\be
h_o = (r - r_+)^{\nu} \,,
\ee
with
\be
2 \nu = -1 + \sqrt{1+\frac{8 \a^2}{\a^2 + 4 (\mu/\gamma)^2}} > 0 \,. \label{eq:nu}
\ee
Here $\gamma$ is a constant that determines the normalization of charge as described in the Appendix.
The exponent $\nu$ is just the dimension of the operator $T^{xy}$ in this emergent IR scaling geometry (more precisely, viewed as a scalar operator in the IR with standard quantization, it has dimension $\delta = 1 + \nu$). It is a well established fact that, at low temperatures and frequencies (here we have $\omega = 0$), the scaling of the imaginary part of the Green's function is directly given by the dimension of the operator in the IR theory. This follows from a simple matching argument, versions of which can be found, for instance, in \cite{Gubser:2008wz, Faulkner:2011tm, Donos:2012ra}. Applying these arguments to equation (\ref{eq:hxyeq}) for $h_o$, and using the horizon formula (\ref{eq:h2}) to obtain $\eta/s$, gives
\be
\frac{\eta}{s} \sim T^{2 \nu} \,, \quad \text{as} \quad T \to 0 \,. \label{eq:result}
\ee
In particular, it is directly the ratio $\eta/s$ that is controlled by the exponent $\nu$.
We have verified this expectation by numerically solving the perturbation equation for $h_o$ and using the formula (\ref{eq:h2}). Plots showing this agreement are given in Appendix \ref{sec:linear}. Note that $2 \geq 2 \nu \geq 0$, so that for all values of the ratio $\alpha/\mu$ the low temperature behavior is consistent with the entropy production bound (\ref{eq:newbound}). When $\alpha = 0$, from (\ref{eq:nu}) we have $\nu = 0$, recovering $\eta/s \sim \text{const.}$ for the extremal Reissner-Nordstr\"om black hole \cite{Edalati:2009bi}. When $\mu = 0$ we have $\nu = 1$ from (\ref{eq:nu}), recovering $\eta/s \sim T^2$, as found for the neutral linear axion backgrounds studied in the previous section.

The above scaling argument generalizes to any emergent IR geometry, with general dynamical critical exponent $z$ \cite{Kachru:2008yh, Hartnoll:2011fn}. For instance $z=1$ corresponds to an emergent $AdS_4$ at zero temperature, of the sort we will see in the following section. A result of the form (\ref{eq:result}) will still hold. In that case, we will find
$\nu = 0$, corresponding to $(\delta g)^x{}_y$ being a massless scalar field in the near horizon region and hence describing a marginal operator. In general, $\eta/s \sim \text{const.}$ at low temperatures, independently of the value of this constant, corresponds to $T^{xy}$ marginal (viewed as a scalar operator) in the low energy fixed point theory.

The expression (\ref{eq:result}) in principle allows for a scaling dimension $\nu > 1$. This would lead to very slow entropy production, in contradiction with the putative bound (\ref{eq:newbound}). We are not aware of a general argument that rules this possibility out. The cases studied in this paper all turn out to have $\nu \leq 1$.

\section{Isotropic Q-lattices}

\subsection{Neutral Q-lattices}

To test the ideas of section \ref{sec:entropy} further, we consider a different class of simple momentum-relaxing spacetimes known as Q-lattices \cite{Donos:2013eha}. The action again involves gravity coupled to scalar fields:
\begin{equation}
S = \frac{1}{16\pi G_N}\int \mathrm{d}^4 x\sqrt{-g}\left\{R+\frac{6}{L^2}-2\sum_{I=1}^2\left[(\nabla_a \Phi_I)(\nabla^a \Phi_I)^\dagger+V_I(|\Phi_I |^2)\right]\right\} \,.
\label{eq:action2}
\end{equation}
Here $\Phi_I$ are two complex scalar fields and $L$ is the AdS radius, which henceforth we will set to one. The equations of motion are
\begin{subequations}
\begin{equation}
R_{ab}+3 g_{ab}=\sum_{I=1}^2\left[(\nabla_a \Phi_I)(\nabla_b \Phi_I)^\dagger+(\nabla_b \Phi_I)(\nabla_a \Phi_I)^\dagger+g_{ab}V_I(|\Phi_I |^2)\right] \,,
\end{equation}
and
\begin{equation}
\Box \Phi_I-V^\prime_I(|\Phi_I|^2)\Phi_I=0\,.
\end{equation}
\end{subequations}
Because we are interested in isotropic systems, we choose $V_1 = V_2 = V$. Furthermore, for simplicity we will take the potential
\begin{equation}
V(\eta) = -2 \eta\,.
\end{equation}

As for the linear axion models above, we are interested in solutions that break translational invariance explicitly in the matter sector, but with a metric that is translationally invariant and isotropic along the boundary directions. Unlike the linear axion case, however, the background itself 
must be found numerically.
We will use the Einstein-DeTurck trick to generate the background solutions. This method was first introduced in the seminal work of \cite{Headrick:2009pv}, further developed in \cite{Figueras:2011va}, and recently reviewed in \cite{Dias:2015nua}. Some details are given in Appendix \ref{sec:appendQ}. The metric takes the form
\be
\mathrm{d} s^2 = -a(r)\mathrm{d}t^2+ b(r) \mathrm{d}r^2+ c(r) \left(\mathrm{d}x_1^2+\mathrm{d}x_2^2 \right)\,, \label{eq:Qmet}
\ee
while translation invariance is broken (isotropically) by the scalars:
\begin{equation}
\Phi_I = \phi(r) e^{i\,k \, x_I} \qquad\text{for}\qquad I\in\{1,2\}\,, \label{eq:PPhi}
\end{equation}
where $\phi$ is real. Note that different coordinates are used in the Appendix; in the main text we use notation similar to that in the linear axion discussion above.

The important physical difference with the linear axion solutions is that translation invariance is now broken by a finite $k$ source (\ref{eq:PPhi}). Finite $k$ perturbations of $AdS_4$ decay exponentially towards the IR. Therefore a simple expectation is that, at least so long as the lattice source is not too large, the zero temperature IR geometry will again be $AdS_4$ rather than the $AdS_2 \times \R^2$ of the linear axion case above. This expectation is verified in Appendix \ref{sec:appendQ} for the numerically constructed Q-lattice geometries. In particular, this means that these neutral solutions do not have a zero temperature ground state entropy density. It also means that translation invariance is restored in the far IR.

The isotropic, neutral Q-lattice solutions we construct in this way depend on three parameters: wavevector $k$, temperature $T$ and strength $V$ of the lattice. $V$ is the boundary value of $\phi(r)$ in (\ref{eq:PPhi}), defined more carefully in the Appendix. Scale invariance of the microscopic boundary theory means that physical quantities can only depend on the ratios $\{k/V,T/V\}$. We are interested in low temperatures $T/V\ll1$, with $k/V$ kept fixed.

\begin{figure}[h]
\centering
\includegraphics[height = 0.35\textheight]{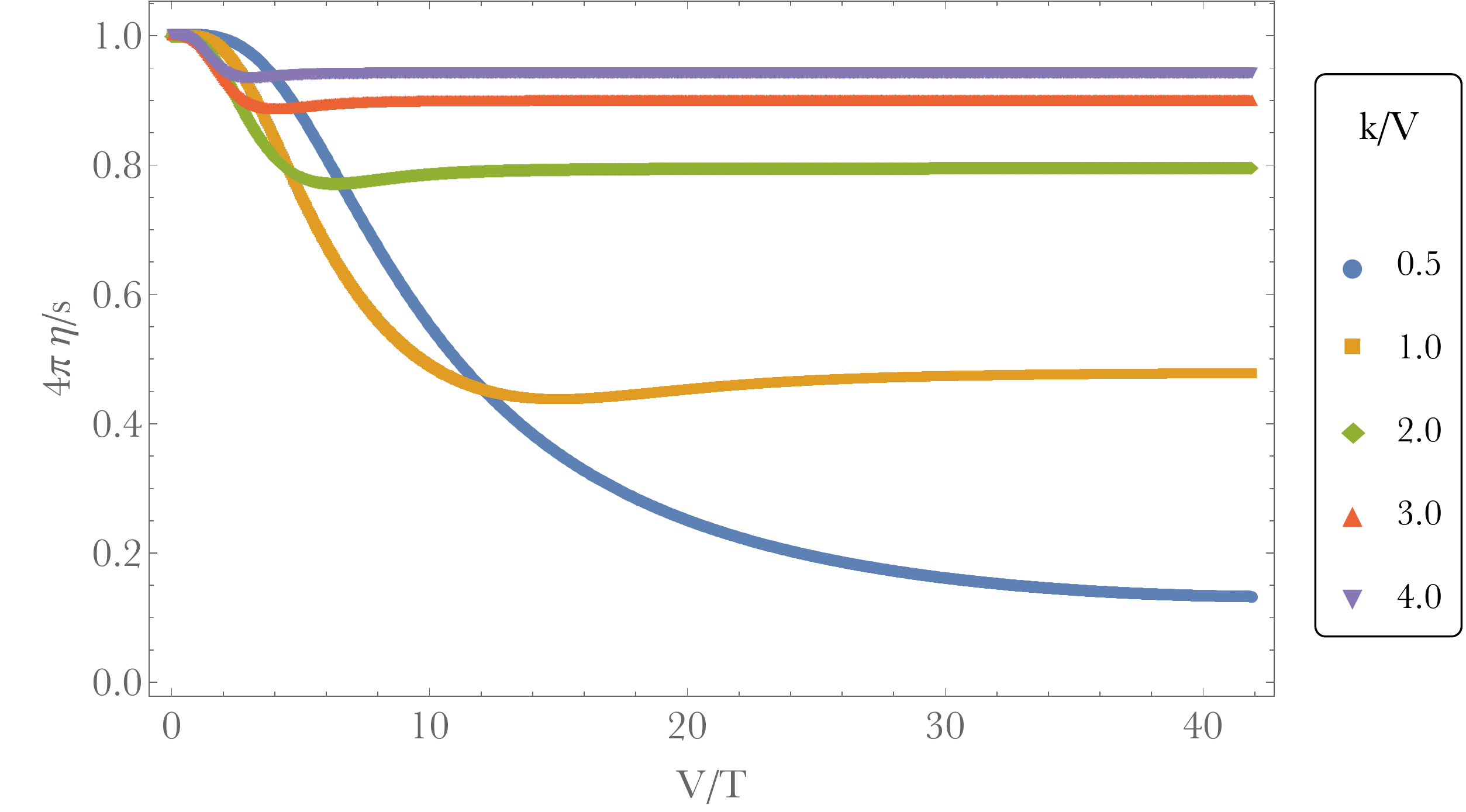}
\caption{\label{fig:qlattices}  {\bf Plot of $4\pi \, \eta/s$ as a function of $V/T$} for several values of $k/V$, for neutral Q-lattice backgrounds. As anticipated, $\eta/s < 1/(4 \pi)$ at all finite temperatures, and $\eta/s \to \text{const.}$ as $T \to 0$.}
\end{figure}

With the numerically constructed backgrounds at hand, the viscosity over entropy density ratio is obtained by perturbing the background as described in section \ref{sec:formula}, and using equation (\ref{eq:h2}). Some details are given in Appendix \ref{sec:appendQ}. The mass squared of the metric perturbation $\left(\delta g\right)^x{}_y$ is seen to be positive everywhere: $m^2 = 4 k^2 \phi^2/c$ in the notation of (\ref{eq:Qmet}). In particular, it is proportional to the strength of the lattice squared, $\phi(r)^2$. We have explained above that the Q-lattice scalar field decays to zero on the horizon as $T \to 0$. The general arguments in section \ref{sec:formula} and the end of section \ref{sec:scale} therefore imply that we will have $\eta/s < 1/(4 \pi)$ and that $\eta/s$ will asymptote to a finite nonzero value as $T \to 0$. These expectations are borne out by explicit numerical solutions to the perturbation equation, as shown in figure \ref{fig:qlattices}.

We have noted that, in these cases, translation invariance is restored in the far IR (i.e. on the low temperature horizon, as the $\left(\delta g\right)^x{}_y$ mass goes to zero). Therefore the low temperature horizon will be a standard planar Schwarzschild-AdS horizon and will have a `membrane paradigm' $\left. \eta/s \right|_\text{m.p.}= 1/(4\pi)$ \cite{Iqbal:2008by}. However, because these components of the graviton are massive at intermediate radii in the bulk, this membrane paradigm $\eta/s$ is no longer equal to the boundary field theory $\eta/s$. Figure \ref{fig:qlattices} shows that the constant low temperature value of $\eta/s$ is becoming parametrically small as $V/k$ becomes large. To understand this properly, we now turn to the zero temperature dependence of $\eta/s$ on $V/k$.

The results described above suggest that the isotropic Q-lattice solutions tend to $AdS_4$ in the far IR. Motivated by this fact, we construct novel zero temperature solutions, which interpolate between two $AdS_4$ geometries (in the UV and IR). These should be the zero temperature limit of the backgrounds we have been considering so far. The construction of these solutions is in many respects similar to what we have done before. For completeness we present the metric and scalar field \emph{ans\"atze} in Appendix \ref{sec:appendQ}, as well as the respective reference metric.

At zero temperature, one can use perturbation theory in $V$ to predict the behavior of $\eta/s$. The method is very similar in spirit to that used in \cite{Hartnoll:2014gaa} so we will just quote the final result up to tenth order in $V/k$:
\begin{multline}
4\pi  \frac{\eta}{s} =1-\left(\frac{V}{k}\right)^2+\frac{463}{512} \left(\frac{V}{k}\right)^4-\frac{1594307}{1990656} \left(\frac{V}{k}\right)^6+\\
\frac{92063924633}{130459631616} \left(\frac{V}{k}\right)^8-\frac{11402485082909330183}{18345885696000000000} \left(\frac{V}{k}\right)^{10}+\mathcal{O}\left[\left(\frac{V}{k}\right)^{12}\right]\,. \label{eq:Qpert}
\end{multline}
Here and in the zero temperature plot below, $\eta/s$ is computed using the horizon formula (\ref{eq:h2}). For an emergent $AdS_4$ horizon, which has vanishing horizon area, this requires taking $\omega \to 0$ first with $T/V$ fixed, and then taking $T/V \to 0$. This is the $T \to 0$ limit appropriate for comparison with the nonzero temperature plots of figure \ref{fig:qlattices}.

\begin{figure}[h]
\centering
\includegraphics[height = 0.3\textheight]{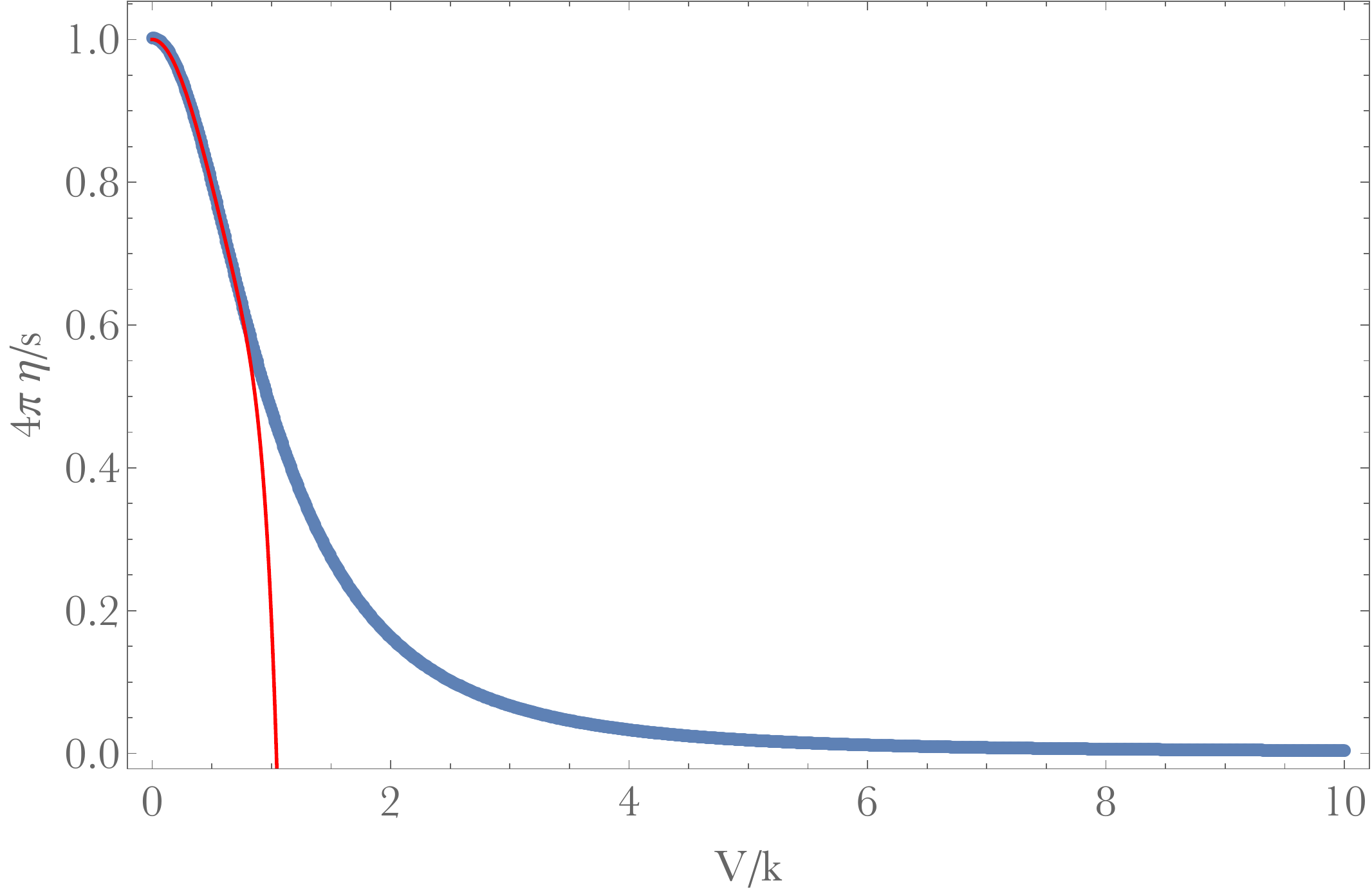}
\caption{\label{fig:zeroeta}  {\bf Plot of $4\pi \eta/s$ as a function of $V/k$ at zero temperature}, for the isotropic, neutral Q-lattices. The blue dots are numerical data points, whereas the red line is the analytic perturbative result (\ref{eq:Qpert})}
\end{figure}

Numerically, one can compute $\eta/s$ directly at zero temperature, and for all values of $V/k$, inclusive at large $V/k$. This is depicted in Fig.~\ref{fig:zeroeta}, where the disks represent the numerical data, and the solid red line the perturbative result (\ref{eq:Qpert}). The agreement between perturbation theory and the numerical results is reassuring.
The strong suppression of $\eta/s$ at large $V/k$ is straightforward to understand mathematically. $V$ sets the magnitude of the scalar field $\phi$ and hence also of the mass squared $m^2 \sim \phi^2$. At large $V/k$, the mass squared becomes very large in the interior of the spacetime, while vanishing towards the boundary and the Poincar\'e horizon. The imaginary, dissipative, part of the two point function -- that determines the shear viscosity -- is given by the probability that a perturbation $\left(\delta g\right)^x{}_y$ can tunnel from the boundary through to the horizon (an early use of this general fact is in \cite{Son:2002sd}). When the mass squared is large in the interior, the tunneling rate is small (probably exponentially so at large $V/k$, although we do not have a wide enough data range for a clean fit) and hence so is $\eta/s$.

The suppression seen in figure \ref{fig:zeroeta} can also be understood physically, as can the presence of a finite nonzero $\eta/s < 1/(4 \pi)$ as $T \to 0$ in these models. The neutral Q-lattices exhibit a restoration of translation invariance in the far IR of the extremal or near-extremal geometries. This means that, in fact, at low temperatures there will exist a long lived hydrodynamic mode, corresponding to the emergent conserved momentum in the IR fixed point theory. The mode has a decay rate $\Gamma$ at any $T>0$, but the decay rate goes to zero as $T \to 0$. This decay rate can be obtained by a combination of memory matrix methods and holography \cite{Hartnoll:2012rj}. Because the Q-lattice decays exponentially (due to being a finite $k$ mode) in $AdS_4$, the decay rate will be exponentially small $\Gamma \sim e^{-1/T}$ \cite{Hartnoll:2012rj}. Thus at low temperatures, the emergent momentum will be conserved to a very good approximation. Over a range of wavevectors, the emergent momentum density will therefore diffuse (the crossover between diffusion and exponential decay in this kind of situation is discussed in detail in \cite{Davison:2014lua}). This diffusivity will define an IR shear viscosity. The IR shear viscosity is a property of the near horizon Schwarzschild-AdS geometry and therefore will in fact be $\left. \eta/s \right|_\text{IR} = \left. \eta/s \right|_\text{m.p.} = 1/(4\pi)$. This is a physical diffusive mode in this system. However, at large $V/k$, momentum is strongly non-conserved at intermediate energy scales, and so this mode has a very small overlap with the UV energy momentum tensor operator that we are using in the definition of $\eta$ in (\ref{eq:eta}). The constant $\eta/s$ at zero temperature plotted in figure \ref{fig:zeroeta} is a measure of the overlap of the energy momentum tensor operator in the UV theory with the emergent conserved momentum operator in the IR theory. This overlap can be defined precisely as a thermodynamic susceptibility, as discussed in \cite{Hartnoll:2012rj}.

\begin{figure}[h]
\centering
\includegraphics[height = 0.3\textheight]{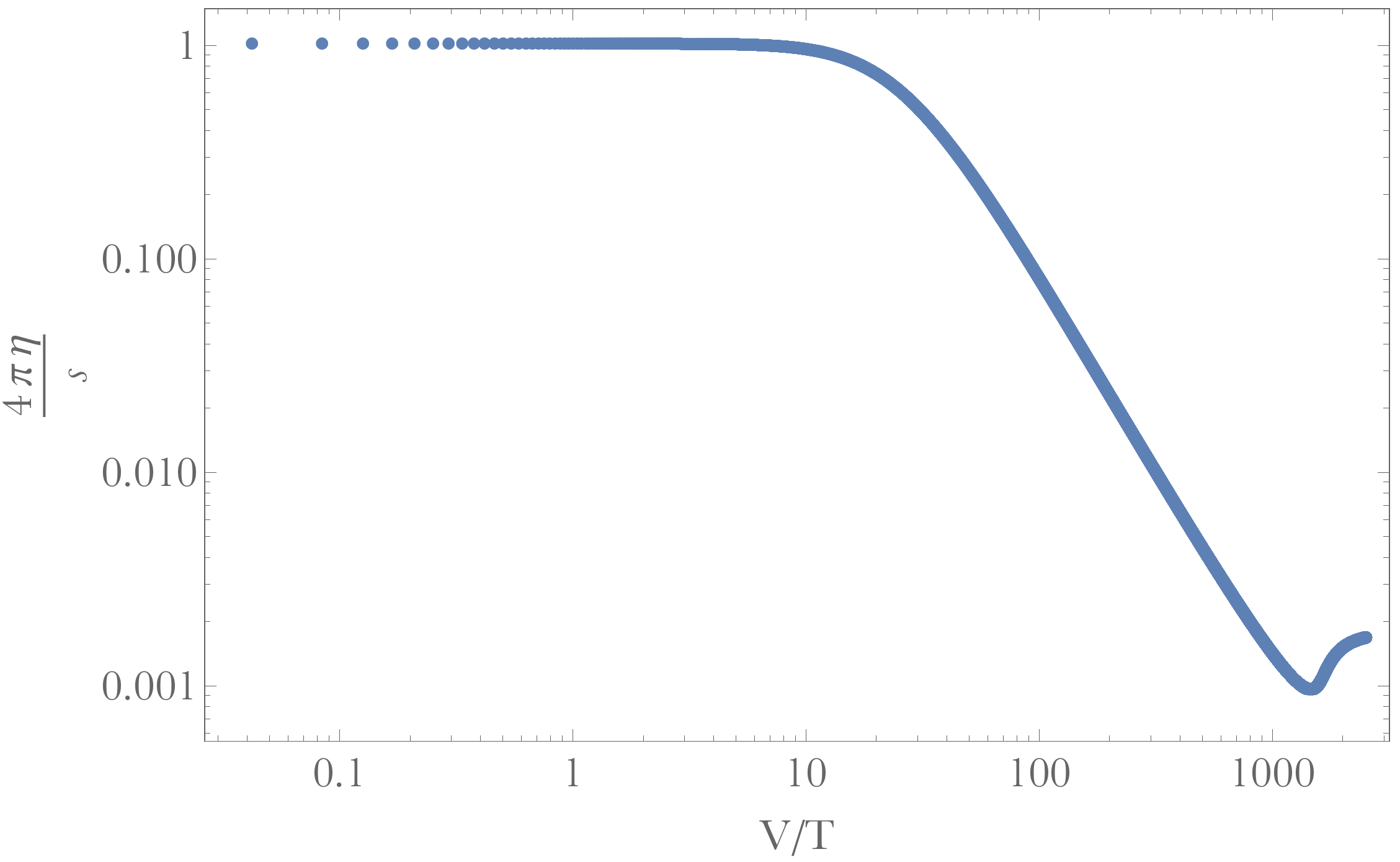}
\caption{\label{fig:intermediate2}  {\bf Log-log plot of $4\pi \eta/s$ as a function of $V/T$}, for the isotropic, neutral Q-lattice with $V/k = 10$. The intermediate scaling regime has $\eta/s \sim T^2$.}
\end{figure}

Finally, because the zero temperature limit of the ratio $\eta/s$ is very small in figure \ref{fig:zeroeta}, one should ask how quickly these small values are approached as the temperature is lowered. Figure \ref{fig:intermediate2} shows that, for a large value of $V/k$, an intermediate temperature regime scaling like $\eta/s \sim T^2$ is clearly discernible.
This is the region of fastest decrease, and therefore the approach to the small constant is consistent with the potential bound (\ref{eq:newbound}). It is likely that the intermediate $T^2$ behavior here is related to the $T^2$ seen for the neutral linear axion model in equation (\ref{eq:lowTlinear}) and figure \ref{fig:scalars}: When the wavevector $k$ is small compared to other scales one may be able to effectively linearize the exponential in (\ref{eq:PPhi}).

\subsection{Charge Q-lattices}

We have also considered isotropic charged Q-lattices, first constructed in \cite{Donos:2013eha}. The phenomenology becomes richer, since there are multiple possible zero temperature IR geometries. The action remains (\ref{eq:action2}), except with a kinetic term for a Maxwell field:
\begin{multline}
S = \frac{1}{16\pi G_N}\int \mathrm{d}^4 x\sqrt{-g}\left\{R+\frac{6}{L^2}-2\sum_{I=1}^2\left[(\nabla_a \Phi_I)(\nabla^a \Phi_I)^\dagger+V_I(|\Phi_I |^2)\right]-\frac{1}{\gamma^2} F^2\right\}\,,
\label{eq:action3}
\end{multline}
where $F=\mathrm{d}A$ is the Maxwell field strength. We will set $L = 1$. On the charged Q-lattice solutions, the metric and scalar field have the same form as in the neutral case, (\ref{eq:Qmet}) and (\ref{eq:PPhi}) respectively. The Maxwell gauge field is then $A = A_t(r)\mathrm{d}t$ with $\mu = \lim_{r\to+\infty} A_t(r)$ defining the chemical potential.

The background solutions depend on three dimensionless ratios $\{k/V,\mu/(\gamma V),T/V\}$, which makes an exhaustive study more difficult. We again use the Einstein-DeTurck trick to solve for the backgrounds, and compute the resulting $\eta/s$. However, as always, we can make educated guesses about the zero temperature IR geometry, and hence the low temperature behavior of $\eta/s$, prior to solving the full equations of motion.

The analysis of \cite{Donos:2013eha} shows that one should expect at least three distinct IR geometries depending on the values of $\gamma k/\mu$. This is done by starting with a zero temperature near-horizon $AdS_2 \times \R^2$ geometry -- this could be the IR of an extremal Reissner-Nordstr\"om black hole, but also of any other solution in which the lattice has become small close to the horizon. The $AdS_2 \times \R^2$ background is given by
\begin{equation}
\mathrm{d}s^2_{IR} =\frac{1}{6} \left(-\rho^2\,a_0^2\,\mathrm{d}t^2+\frac{\mathrm{d}\rho^2}{\rho^2}\right)+b_0^2 \left(\mathrm{d}x_1^2+\mathrm{d}x_2^2 \right)\quad \text{and}\quad A = \frac{a_0}{2\sqrt{3}} \, \rho \, \mathrm{d}t \,. \label{eq:nh}
\end{equation}
Here the AdS$_2$ extremal horizon is at $\rho=0$ and $\{a_0,b_0\}$ are two real constants that account for how time and distances are measured in the IR. These can only be fixed with a full solution that interpolates from the UV to the IR.

In the near-horizon background (\ref{eq:nh}), the magnitude of the Q-lattice in (\ref{eq:PPhi}) behaves as
\begin{equation}
\phi = \rho^{\theta} \,,
\end{equation}
with exponent
\begin{equation}
\theta=-\frac{1}{2} + \sqrt{\frac{k^2}{6b_0^2}-\frac{1}{12}}\,.
\end{equation}
For small enough $V$, we can obtain $b_0$ from the metric of a planar extremal Reissner-Nordstr\"om black hole. It turns out to be $b_0 = \mu/(\g \sqrt{3})$, and so one finds
\begin{equation}
\theta=-\frac{1}{2} + \sqrt{\frac{\g^2 k^2}{2\mu^2}-\frac{1}{12}}\,.
\end{equation}
It is now easy to identify three regimes for $\g k/\mu$.
\begin{itemize}
\item If $\g k/\mu>\sqrt{2/3}$, then $\theta > 0$ and the lattice is irrelevant in the IR. Translation invariance is restored and hence we expect $\eta/s$ to tend to a constant at low temperatures.
\item If $1/\sqrt{6}< \g k/\mu <\sqrt{2/3}$, then $\theta < 0$ and the lattice is relevant. The new IR geometry is likely an insulator, in the sense of \cite{Donos:2012js}. If, as sometimes occurs, there is an emergent scaling symmetry in the insulating phase (in particular, it is not gapped), then, because the lattice is supporting the IR, we expect $\eta/s$ to go to zero like a power of $T$ at low temperatures, analogously to the linear axion models considered above. More generally, there will be a complicated $T$ dependence.
\item If $\g k/\mu <1/\sqrt{6}$, then $\theta$ is complex, giving an instability of $AdS_2 \times \R^2$. The new IR geometry will be a close cousin of the ones presented in \cite{Horowitz:2009ij}. The ratio $\eta/s$ will be temperature dependent at the lowest temperatures; there is no guarantee it will be a simple power law.
\end{itemize}
The above analysis, of course, neglects possible phase transitions associated with large values of $V/\mu$ as well as possible first order phase transitions.

\begin{figure}[h]
\centering
\includegraphics[height = 0.3\textheight]{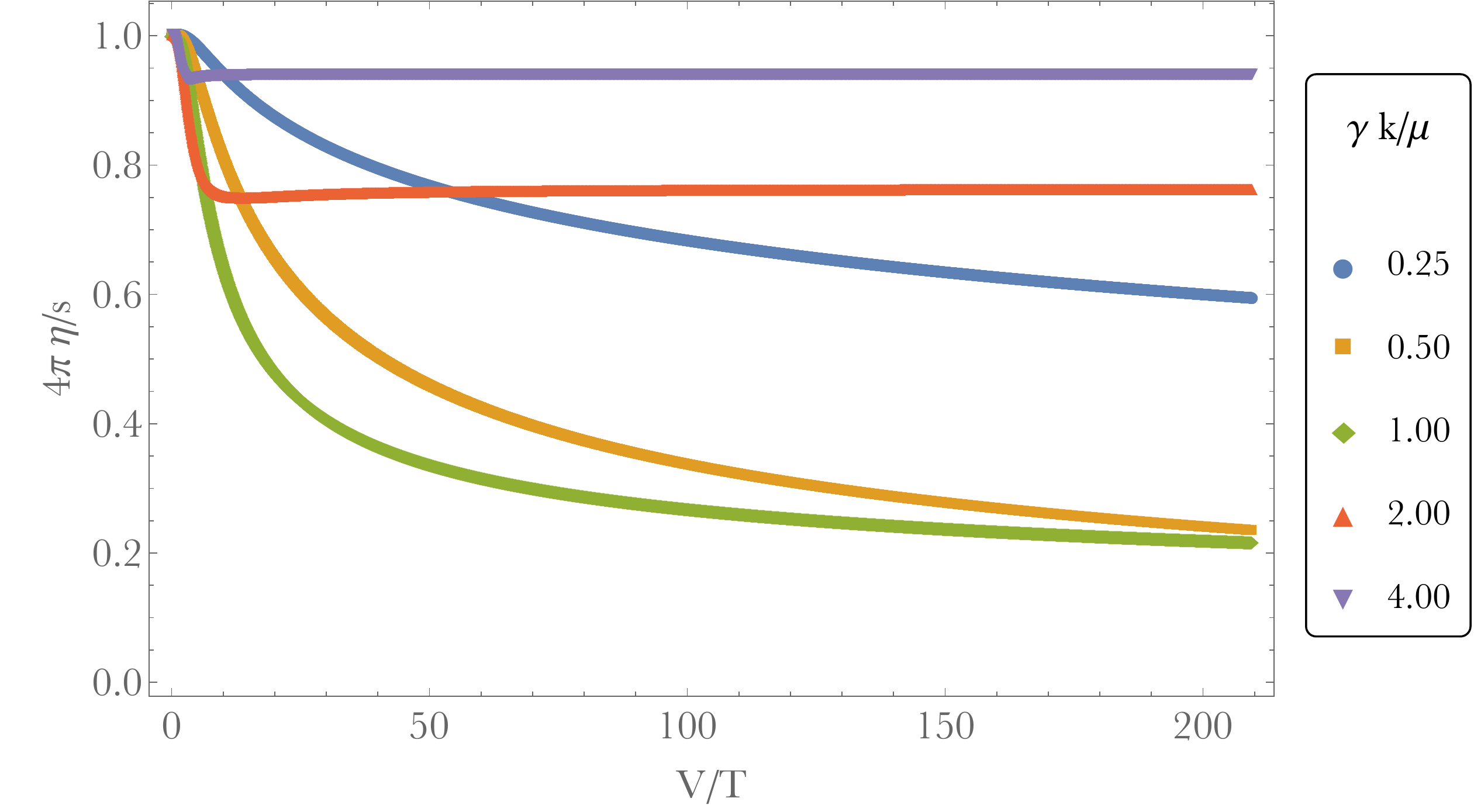}
\caption{\label{fig:charged}  {\bf Plot of $4\pi \eta/s$ as a function of $V/T$} for several values of $\gamma k/\mu$ and with $\mu/(\gamma V)=1$, for the isotropic, charged Q-lattices.}
\end{figure}

The numerical construction of the charged Q-lattices is very similar to the neutral case. The equation that determines $\eta/s$ has the same mass term, except that the scalar field profile is now determined via the equations of motion derived from (\ref{eq:action3}). We have probed the three different regimes of $\gamma k/\mu$, and the results are plotted in Fig.~\ref{fig:charged}. We have tried other values of $\mu/V$, but the curves look qualitatively the same.
For the larger three values of $\g k/V$, we indeed see that $\eta/s$ is tending to a constant, as anticipated for cases with an irrelevant lattice and restored translation invariance. At larger lattice strength, the constant is smaller, as in the neutral case. For small values of the lattice strength $(\g V)/\mu$ and for $\g k/\mu>\sqrt{2/3}$ the constant reached at zero temperature can be predicted using perturbation theory (see appendix \ref{ap:perturbads2} for more details). Because the zero temperature IR is $AdS_2 \times \R^2$ here, rather than the $AdS_4$ of the neutral case considered above, the lattice dies off as a power law rather than exponentially towards the horizon \cite{Hartnoll:2012rj}.

The case that is most novel compared to those that we have discussed so far is the insulating regime. Translation invariance is strongly broken in the IR. It is difficult to extract a clear asymptotic low temperature behavior of $\eta/s$ in this case ($\gamma k/\mu = 0.5$ in figure \ref{fig:charged}). We have run the numerics down to considerably lower temperatures. Over the temperature range we have probed, the behavior is close, but not exactly equal to, a power law that is considerably weaker than $T^2$. A power law or not, however, the behavior is weaker than the putative bound (\ref{eq:newbound}). Finally, $\gamma k/\mu = 0.25$ in figure \ref{fig:charged} is in the range where the scalar field has complex scaling dimensions from the perspective of the would-be AdS$_2$. In this regime, $\eta/s$ also decreases slowly, but it is inconclusive -- over the temperature range we have probed -- whether it will reach zero or if it will reach a constant.

\section{Discussion}

In this paper we have focused on simple holographic models for breaking translation invariance (i.e. ones that do not require solving PDEs), with a view to probing the rate of entropy production and possible bounds on $\eta/s$. While we have considered various qualitatively different cases, we have certainly not been exhaustive. For instance, another simple model that breaks translation invariance is `massive gravity' in the bulk \cite{Vegh:2013sk}. This model has uncertain microscopic foundations in general, and so we have not studied it in detail.
However, as a check of the entropy production bound (\ref{eq:newbound}), we can quickly obtain the low temperature behavior of the viscosity over entropy density for the simplest instance of this class of theories ($\beta = 0$ in the terminology of \cite{Vegh:2013sk}). In the neutral theory, the extremal near horizon geometry is $AdS_2 \times \R^2$, and the mass for $\left(\delta g\right)^x{}_y$ is found to be nonzero on the extremal horizon. An analysis very much like that in section \ref{sec:scale} above for the linear axion model gives
\be
\frac{\eta}{s} \sim T^{\sqrt{5} - 1} \sim T^{1.24} > T^2 \qquad \text{as} \qquad T \to 0 \,,
\ee
consistently with (\ref{eq:newbound}). In Appendix \ref{sec:massive} we give some details of the calculation leading to this result.

Several lessons we have found will likely continue to hold in more generic scenarios, in which the bulk spacetime itself is inhomogeneous. For instance, whenever translational symmetry breaking is irrelevant at low energy scales, we expect $\eta/s$ to tend to a constant at low temperatures. This constant will become very small as translation symmetry breaking becomes strong. In constrast, when translational symmetry breaking survives to the lowest energy scales, we expect $\eta/s$ to go to zero at low temperatures like some power of $T$, with the power depending on operator dimensions at the low energy fixed point theory.

Entropy production is a central aspect of physical processes. The boundedness of entropy production has appeared in several different contexts. For instance, the Boltzmann equation (which is directly connected to transport quantities such as the viscosity in weakly interacting systems) can be derived by extremizing entropy production \cite{ziman}. The appropriate quantum observable to think about may be entanglement entropy, whose production is also subject to bounds, e.g.  \cite{bounds, Hartman:2015apr}. A different class of bounds on entanglement entropy leads to important constraints on the scale dependence of quantum dynamics, e.g. \cite{Casini:2012ei, Casini:2015woa}.
A seemingly unrelated type of bound are those on operator dimensions in CFTs, e.g. \cite{Rattazzi:2008pe}. However, operator dimensions control the time dependence of correlators and hence are also directly related to the rate of entropy production -- as indeed we have seen in several instances in this paper. The time may be ripe for this circle of ideas to lead to a possible `meta-theory' of bounds that would, among other things, conclusively prove whether or not physical quantities like $\eta/s$ are bounded in some interesting sense.

\section*{Acknowledgements}

We are grateful to Blaise Gouteraux for helpful comments.
SAH is partially supported by a DOE Early Career Award and the Templeton foundation.

\section*{Note added}

Some of the results presented in this paper have also been independently obtained by \cite{Alberte:2016xja, Burikham:2016roo}, which appeared in parallel.

\appendix

\section{Solving the perturbation equation: linear axion model}
\label{sec:linear}

Equation (\ref{eq:linear}) for the metric perturbation becomes particularly simple if we perform a change of coordinates $z=r_+/r$, which gives
\begin{equation}
z^2 \left[2-\beta ^2 z^2-(2-\beta ^2) z^3\right] h^{\prime\prime}-z \left[4+z^3 (2-\beta ^2)\right]h^\prime-2z^2\beta^2 h=0\,.
\label{eq:scalarsh}
\end{equation}
Here ${}^\prime$ denotes differentiation with respect to $z$ and $\beta \equiv \alpha/r_+$. As anticipated from the underlying conformal symmetry of the UV theory, all physical quantities can only depend on $\beta^2$. In this appendix we drop the subscript, so that $h_o \to h$.

\subsection{Second order perturbation theory in \texorpdfstring{$\beta^2$}{test}}

It is fairly easy to compute $\eta/s$ perturbatively in $\beta^2$. We first expand the function $h$ as
\begin{equation}
h(z) = \sum_{i=0}^{+\infty}\beta^{2i}\,h_{2i}(z)\,
\end{equation}
and expand (\ref{eq:scalarsh}) order by order in $\beta^2$. At first order one finds
\begin{equation}
z \left(1-z^3\right) h^{\prime\prime}_0-\left(2+z^3\right)h^\prime_0=0\,.
\end{equation}
The only solution compatible with our boundary conditions is given by $h_0(z)=1$. To second order in $\beta^2$ one finds
\begin{equation}
z \left(1-z^3\right) h^{\prime\prime}_2-\left(2+z^3\right)h^\prime_2-z=0\,.
\end{equation}
for which the relevant regular solution compatible with our boundary conditions reads
\begin{equation}
h_2(z) = \frac{1}{\sqrt{3}}\arctan\left(\frac{\sqrt{3} z}{z+2}\right)-\frac{1}{2} \log \left(1+z+z^2\right)\,.
\end{equation}
At fourth order one finds
\begin{equation}
z \left(1-z^3\right) h^{\prime\prime}_4-\left(2+z^3\right)h^\prime_4-z\left[\frac{1}{\sqrt{3}}\arctan\left(\frac{\sqrt{3} z}{z+2}\right)-\frac{1}{2}\log \left(1+z+z^2\right)-\frac{z^2 (1-2 z-2 z^2)}{2 \left(1+z+z^2\right)^2}\right]\,.
\end{equation}
The solution with the relevant boundary condition reads
\begin{equation}
h_4(z) = \frac{5}{6 \sqrt{3}} \arctan\left(\frac{\sqrt{3} z}{2+z}\right)-\frac{1}{4} \log \left(1+z+z^2\right)-\frac{(1-z) z}{6 \left(1+z+z^2\right)}+I_1(z) \,,
\end{equation}
where
\begin{multline}
I_1(z)\equiv\frac{1}{6}\int_0^z\mathrm{d}w \frac{w}{\left(1-w^3\right)}\Big[w \left(\sqrt{3} \pi -3\log 3\right)-2 \sqrt{3} (1+2 w) \arctan\left(\frac{\sqrt{3} w}{2+w}\right)\\+3 \log \left(1+w+w^2\right)\Big]\,.
\end{multline}
The integral above can be computed for general values of $z$ in terms of dilogarithmic functions evaluated at compex arguments, but we will not need it in what follows. The only important thing to note is that the integral can be done in terms of much simpler functions if $z=1$. In particular, for $z=1$ one finds
\begin{equation}
I_1(1) = \frac{\log ^2 3}{4}-\frac{\pi  \log 3}{3 \sqrt{3}}-\frac{7 \pi ^2}{54}+\frac{1}{6}\psi ^{(1)}\left(\frac{1}{3}\right)
\end{equation}
where $\psi^{(1)}(x)$ is the first derivative of the digamma function (also known as trigamma function).

Using (\ref{eq:h2}), and the definition of $T$ in (\ref{eq:temp}), one finds the following result:
\begin{multline}
\frac{4\pi\,\eta}{s} =1+\frac{\sqrt{3}}{16 \pi }\left(1-\frac{3 \sqrt{3} \log 3}{\pi }\right)\left(\frac{\alpha }{T}\right)^2+\\
\frac{9 \sqrt{3}}{512 \pi ^3}\left[1-\frac{3 \sqrt{3} \pi }{2}-5 \log 3-\frac{\sqrt{3} \log 3}{\pi }+\frac{9 \sqrt{3} \log ^2 3}{2 \pi }+\frac{2 \sqrt{3}}{\pi} \psi^{(1)}\left(\frac{1}{3}\right)\right]\left(\frac{\alpha }{T}\right)^4 +\mathcal{O}\left[\left(\frac{\alpha }{T}\right)^6\right]
\end{multline}
This formula has been quoted in (\ref{eq:highTlinear}) in the main text.

\subsection{Perturbative results close to extremality \texorpdfstring{$\beta^2 = 6$}{test}}

We expand $h(z)$ in a power series around $\beta^2-6$ as
\begin{equation}
h(z) = \sum_{i=0}^{+\infty}(\beta^2-6)^i\,h_i(z)\,.
\end{equation}
The procedure follows \emph{mutatis mutandis} the calculation for small $\beta^2$, so we just quote the intermediate results.

At zeroth order one finds
\begin{equation}
h_0(z) =\frac{1}{(1-z)^2}\left[1-2 z \left(1+z+\frac{1-3 z}{\frac{4}{3}-\log 3}\right)+\frac{\left(1-2 z-2 z^2\right) \log (1+2 z)}{\frac{4}{3}-\log 3}\right]\,,
\end{equation}
while at first order one finds:
\begin{equation}
h_1(z)=-\frac{z^2}{2 (1-z)^3 (1+2z)} \left[2 z \left(1-\frac{5+z}{4-3 \log 3}\right)+\frac{3 (1+2z) \log (1+2z)}{4-3 \log 3}+1\right]\,.
\end{equation}

The results at second order are already too cumbersome to be presented here, but they can be expressed in closed form in terms of polylogarithmic functions. For $\eta/s$ one finds up to this order in perturbation theory:
\begin{equation}
\frac{4\pi\eta}{s} = \frac{512\pi^2}{2187\left(\frac{4}{3}-\log3\right)^2}\left(\frac{T}{\alpha}\right)^2+\mathcal{O}\left[\left(\frac{T}{\alpha}\right)^4\right]\,.
\end{equation}
Note that the term proportional to $(T/\alpha)^3$ is zero. This formula has been quoted in (\ref{eq:lowTlinear}) in the main text.

\subsection{Sef-dual point: \texorpdfstring{$\beta^2 = 2$}{test}}

There is a value of $\beta$ for which (\ref{eq:scalarsh}) can be solved exactly. This was first noted in \cite{Davison:2014lua}, and occurs at $\beta^2 =2$. The solution with the appropriate boundary conditions is given in terms of the Gaussian hypergeometric function:
\begin{equation}
h(z) = \frac{2}{\sqrt{\pi}} \left|\Gamma \left(\frac{5}{4}-\frac{i \sqrt{7}}{4}\right)\right|^2 \, _2F_1\left(-\frac{1}{4}-\frac{i \sqrt{7}}{4},-\frac{1}{4}+\frac{i \sqrt{7}}{4};1;1-z^2\right)\,,
\end{equation}
from which we can read using (\ref{eq:h2})
\begin{equation}
\frac{4\pi\,\eta}{s}=\frac{4}{\pi} \left| \Gamma \left(\frac{5}{4}-\frac{i \sqrt{7}}{4}\right)\right| ^4\,.
\end{equation}
This expression was given in \cite{Davison:2014lua} and is quoted in (\ref{eq:selfduallinear}).

\subsection{Numerical results for any value of \texorpdfstring{$\beta$}{test}}

Of course, equation (\ref{eq:scalarsh}) can be solved for any value of $\beta$ using numerical methods. In order to proceed, we need to specify boundary conditions. At the horizon, located at $z=1$, we demand regularity, which in turns yields
\begin{equation}
h^\prime(1)+\frac{2\beta^2}{6-\beta^2}h(1)=0\,.
\end{equation}
At the boundary, as usual, we demand $h(0)=1$. The equation was discretized using a Chebyshev collocation grid with no less than $200$ points on the grid and using octuple precision. The results
are shown in figure \ref{fig:scalars} in the main text.

\subsection{Nonzero charge density}

The action with a Maxwell field is now
\be
S =\frac{1}{16\pi G_N} \int \mathrm{d}^4 x\sqrt{-g}\left[  \left( R+\frac{6}{L^2}-\frac{1}{2} \left(\nabla_a \vec{\phi}\right)\cdot \left(\nabla^a \vec{\phi}\right) \right) - \frac{1}{\g^2} F^2 \right] \,.
\ee
The isotropic black brane backgrounds with nonzero charge density are
  \begin{align}
    \dd s^2 ={}& {-} f(r) \dd t^2 + \frac{\dd
      r^2}{f(r)} + r^2 \left( \dd x_1^2 + \dd x_2^2 \right)\, ,  \\
    f(r) ={}& r^2 - \frac{\alpha^2}{2} - \left(r_+^2 -
      \frac{\alpha^2}{2} + \frac{\mu^2}{\gamma^2} \right)
    \left(\frac{r_+}{r} \right) + \frac{\mu^2}{\gamma^2}
    \left(\frac{r_+}{r} \right)^2 \, ,  \\
    A_t(r) ={}& \mu \left( 1- \frac{r_+}{r} \right)\, , \\
    \phi^i ={}& \alpha x_i\, , \quad \text{ for } i \in
    \{1,2\}\, .
  \end{align}
As for the
  uncharged solutions discussed in section \ref{sec:branes}, these backgrounds preserve isotropy but break spatial translation invariance when $\alpha \neq 0$. They clearly
  reduce to the previous case when $\mu = 0$. The horizon is located
  at $r=r_+$ and has a temperature of:
  \begin{equation}
    4 \pi T = 3 r_+ - \frac{1}{r_+}\left( \frac{\alpha^2}{2} +
      \frac{\mu^2}{\gamma^2} \right)\, .
  \end{equation}

  As in the neutral case discussed in the main text, the linearized equation of motion for a
  shear perturbation $h^x{}_y(r)$ is that of a free scalar with a
  positive, radially dependent mass:
  \begin{equation}
    \frac{1}{\sqrt{{-} g}} \partial_r \left( \sqrt{{-} g}
      g^{rr} \partial_r h^x{}_y \right) = \frac{\alpha^2
    }{r^2} h^x{}_y\, .
  \end{equation}
  In particular, all of the arguments of section \ref{sec:formula} apply directly to
  these backgrounds, and $\eta/s$ is explicitly determined by the
  regular solution to this wave equation. The equation cannot
  be solved analytically in general, but it is easily solved
  numerically. In the main text we obtain the viscosity to entropy density ratio analytically at low temperatures from this equation. The two plots below show the low temperature dependence of the viscosity over entropy density ratio as well as the agreement of the scaling with the analytic form given in (\ref{eq:nu}) in the main text.
  
  \begin{figure}[h]
\centering
\includegraphics[height = 0.18\textheight]{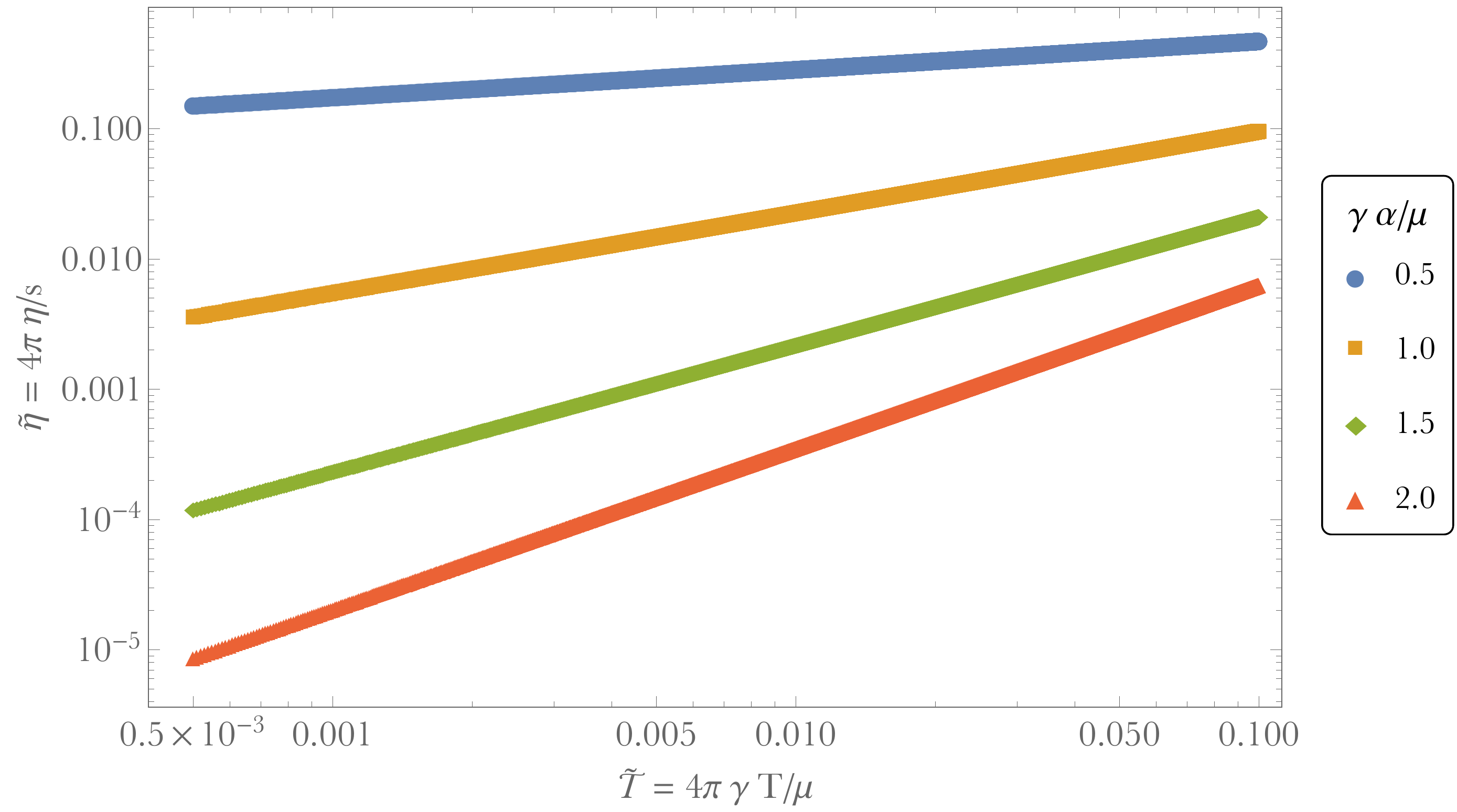}
\includegraphics[height = 0.18\textheight]{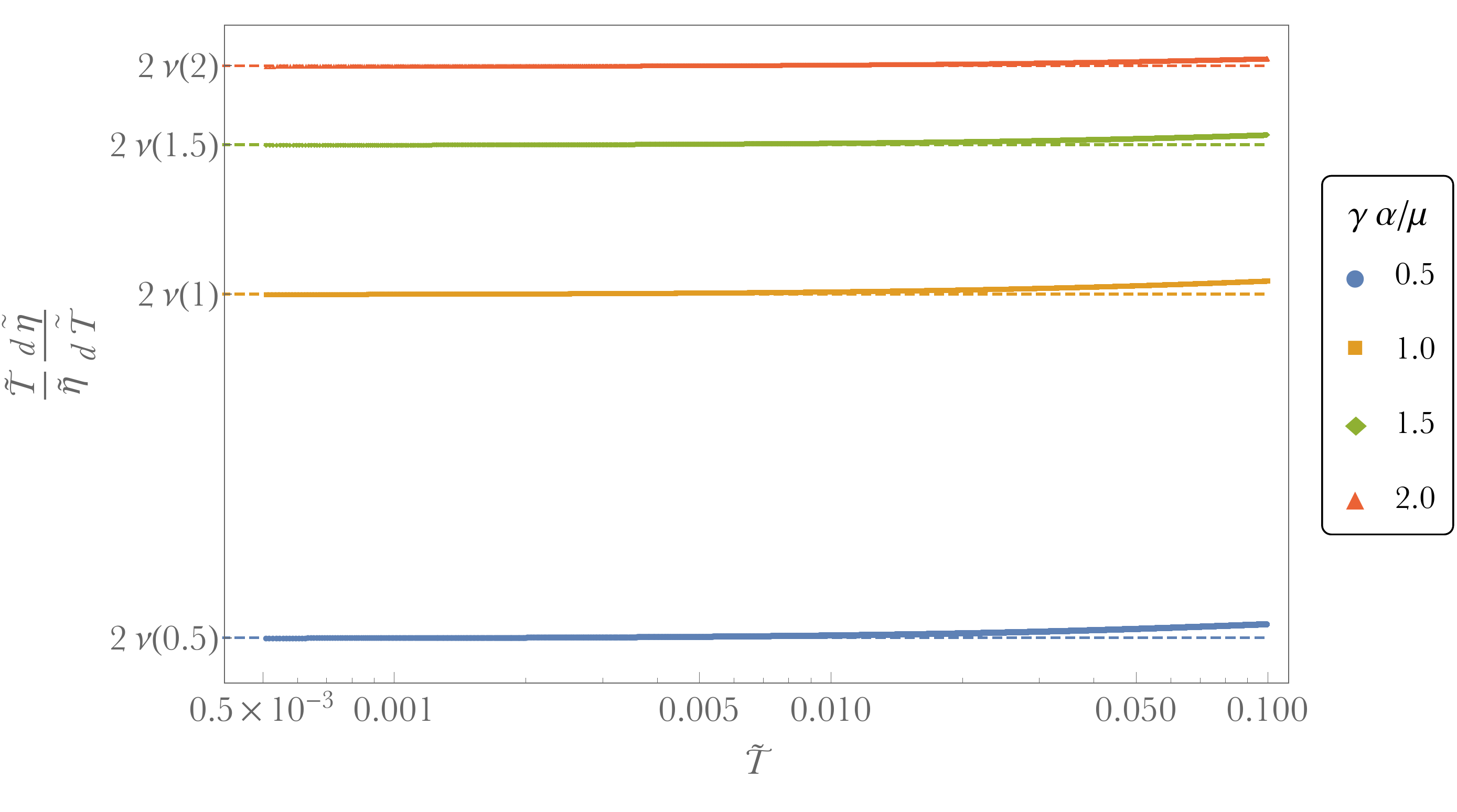}
\caption{\label{fig:scalarscharged}  {\bf Left: Log-log plot of $4\pi \eta /s$ as a function of low $T/\mu$} for different values of $\alpha/\mu$. Power law scaling is seen with the power dependent on $\alpha/\mu$. {\bf Right: Log derivative of $4\pi \eta /s$} with respect to temperature, showing that the powers agree with the analytic prediction of $T^{2\nu}$, with $\nu$ given by (\ref{eq:nu}) in the main text.}
\end{figure}

\section{Technical details for Q-lattices}
\label{sec:appendQ}

We will write the general static, translationally and rotationally invariant spacetime as
\begin{equation}
\mathrm{d}s^2 = \frac{1}{(1-y^2)^2}\left[-A(y)y_+^2 y^2 G(y)\mathrm{d}t^2+\frac{4B(y)\mathrm{d}y^2}{G(y)}+y_+^2 S(y)(\mathrm{d}x_1^2+\mathrm{d}x_2^2)\right]\,,
\end{equation}
with
\begin{equation}
G(y)=3-3y^2+y^4\,.
\end{equation}
Thus we must solve for three functions -- $A,B,S$ -- of a single variable $y$. For the reference metric of the Einstein-DeTurck trick, we will take the line element above with $A=B=S=1$, \emph{i.e.} the four-dimensional planar Schwarzschild black hole. This in turn, fixes the background temperature of our solution to be $T=3 y_+/(4\pi)$. Here, the conformal boundary is located at $y=1$, whereas the horizon is at $y=0$.

Translation invariance is broken by the scalars, which take the form
\begin{equation}
\Phi_I(x_I,y) = (1-y^2)\,e^{i\,k_I x_I} \Psi_I(y)\quad\text{for}\quad I\in\{1,2\}\,,
\end{equation}
where there is no sum over $I$ and the $\Psi_I$ are real. Since we are interested in isotropic Q-lattices, we will take $\Psi_1(y) = \Psi_2(y)=\Psi(y)$ and $k_1 = k_2=k$.

We are thus left with four equations in four unknowns: $\{A,B,S,\Psi\}$. For our choice of scalar field mass, the scalar field $\Psi$ admits the following expansion in Fefferman-Graham coordinates\footnote{The relation between $Z$ and $y$ can be found close to the conformal boundary $y=1$ by solving the equations of motion order by order in $1-y$. It turns out to be $y=1-\frac{y_+}{2}\,Z-\frac{y_+^2}{8}Z^2-\frac{y_+^3}{16}Z^3+\mathcal{O}(Z^4)$.}:
\begin{equation}
\Psi = \Psi^{(1)}Z+\Psi^{(2)}Z^2+\ldots\,.
\end{equation}
We shall use the standard quantization, in which case $\Psi^{(1)}\equiv V$ is the source for an operator dual to $\Phi_I$, whose expectation value is proportional to $\Psi^{(2)}$. Here $V$ has mass dimension $1$. Since we are interested in solutions that asymptote to a flat conformal boundary metric, our parameter space is three-dimensional, depending on $\{V,T,k\}$. Because
the UV theory is conformally invariant, physical quantities can only depend on the dimensionless ratios $\{k/V,T/V\}$.

We discretize the Einstein-Scalar DeTurck equations using a pseudo-spectral collocation method on the Chebyshev-Gauss-Lobatto points, and a Newton-Raphson method to solve the resulting system of algebraic equations.

\begin{figure}[h]
\centering
\includegraphics[height = 0.18\textheight]{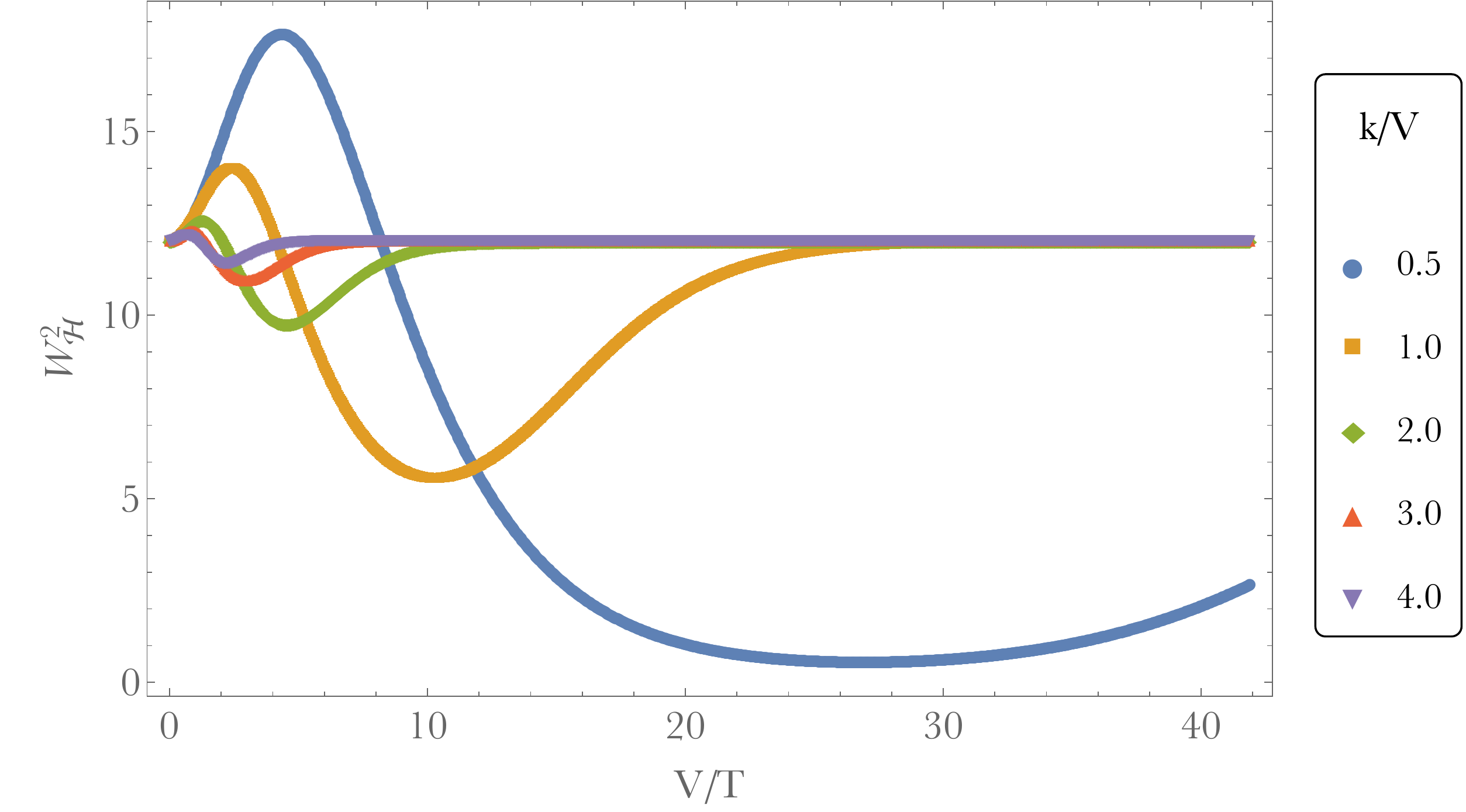}
\includegraphics[height = 0.18\textheight]{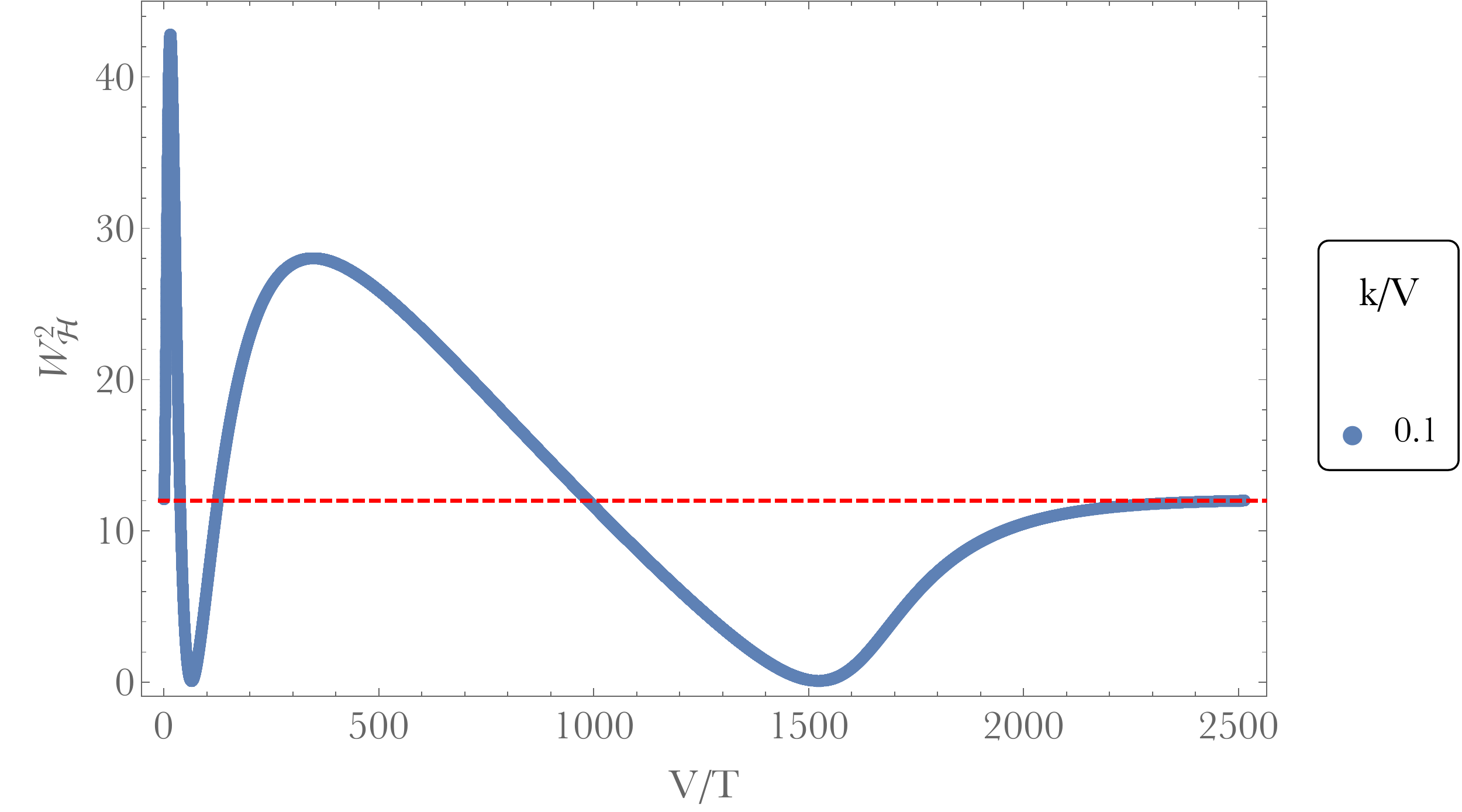}
\caption{\label{fig:background}  {\bf Plots of the Weyl tensor squared at the horizon, as a function of $V/T$} for several values of $k/V$. $W^2$ returns to $12$ at low temperatures, indicating the emergence of an $AdS_4$ spacetime. For small $k/V$, the approach is slow and indirect.}
\end{figure}

In Fig.~\ref{fig:background} we see how the square of the Weyl tensor evaluated at the horizon, $W^2_{\mathcal{H}}\equiv \left.C_{abcd}C^{abcd}\right|_{\mathcal{H}}$, behaves as a function of $V/T$, for several values of $k/V$. In each case we have probed, $W^2_{\mathcal{H}}$ approaches $12$ as $T/V$ is lowered, signalling that the system, as we lower $T$, is returning to planar Schwarzschild at low temperatures, and thus to AdS$_4$ at zero $T$. This gives good evidence that, as anticipated, neutral Q-lattices are irrelevant from the IR perspective. In cases where the lattice strength $V$ is large, the approach to $AdS_4$ at zero temperature can be quite slow and roundabout, as is also shown in Fig.~\ref{fig:background}.

To obtain the shear viscosity, we again perturb the Einstein-scalar equations by
\begin{equation}
\delta g_{x_1 x_2} = g_{x_1 x_1}(y) h_o(y)\,.
\end{equation}
Since the background metric preserves translational symmetry, this perturbation decouples from all other metric and scalar perturbations. The equation for $h_o$ takes a simple form:
\begin{equation}
\Box h_o-\frac{4k^2(1-y^2)^4 \Psi^2}{y_+^2 S}\,h_o=0\,.
\end{equation}
Again, the effective mass squared of Eq.~(\ref{sec:formula}) is manifestly positive, \emph{i.e.} $m(y)^2 = 4k^2(1-y^2)^4 \Psi^2/(y_+^2 S)$. Furthermore, because the scalar field $\Psi$ vanishes on the horizon, so will the mass squared. To complete the system, we need to provide boundary conditions. At the conformal boundary, we impose $h_o(1)=1$, while at the horizon regularity demands $h_o^\prime(0)=0$.

For the zero temperature geometries, the metric and scalar field read:
\begin{subequations}
\begin{equation}
\mathrm{d}s^2 = \frac{1}{y^2}\left[-(1-y)^2 A(y)\mathrm{d}t^2+\frac{B(y)\mathrm{d}y^2}{(1-y)^2}+(1-y)^2S(y)(\mathrm{d}x_1^2+\mathrm{d}x_2^2)\right] \,,
\end{equation}
and
\begin{equation}
\Phi_I(x_I,y) = \frac{y}{1-y}\Psi(y)\,e^{i\,k_I x_I}\quad\text{for}\quad I\in\{1,2\}\,.
\end{equation}
\end{subequations}
The reference metric is obtained by setting $A=B=S=1$, the conformal boundary is located at $y=0$, and the Poincar\'e horizon at $y=1$. At the conformal boundary we demand $A(0)=B(0)=S(0)=1$, $\Psi(0)=V$ and at the Poincar\'e horizon we set $A^\prime(1)=S^\prime(1)=\Psi(1)=0$ and $B(1)=1$.

The computation of $\eta/s$ parallels the finite temperature case, except that the mass term is now changed to $m(y)^2 =4k^2y^4 \bar{\Psi}^2 /[(1-y)^4 S(y)]$.

\subsection{Perturbative expansion about extremal Reissner-Nordstr\"om}
\label{ap:perturbads2}

In figure \ref{fig:charged} in the main text we saw that for large enough $\g k/\mu$, the ratio $\eta/s$ tends to a constant
at zero temperature. In this section we obtain the constant analytically in a perturbation theory in the strength of the lattice $V$. In this appendix we set $\gamma = 1$ for simplicity, it is easily restored.

Write the metric, gauge field and scalar as
\begin{subequations}
\begin{equation}
\mathrm{d}s^2 = -f(r)e^{-2 \delta(r)}\mathrm{d}t^2+\frac{\mathrm{d}r^2}{f(r)}+r^2(\mathrm{d}x_1^2+\mathrm{d}x_2^2)\,,
\end{equation}
\begin{equation}
A = a_t(r)\mathrm{d}t\,,
\end{equation}
and
\begin{equation}
\Phi_I(x_I,r) = e^{i\,k_I x_I} \tilde{\Phi}_I(r)\quad\text{for}\quad I\in\{1,2\}\,,
\end{equation}
\end{subequations}
where there is no sum over $I$ and the $\Psi_I$ are real. Since we are interested in isotropic Q-lattices, we will take $\tilde{\Phi}_1(r) = \tilde{\Phi}_2(r)=\Phi(r)$ and $k_1 = k_2=k$. At zeroth order in perturbation theory, the extremal Reissner-Nordstr\"om black brane is
\bea
f(r) &= & f^{(0)}(r)\;\;\;\equiv\;\;\;  \left(1-\frac{r_+}{r}\right)^2 \left(r^2+2 r_+ r+3 r_+^2\right)\,, \\
 a_t(r) & = & a^{(0)}_t(r)\;\;\;\equiv\;\;\;  \sqrt{3}r_+\left(1-\frac{r_+}{r}\right)\,,\\
 \delta(r) & = &   \Phi(r)\;\;\;=\;\;\;  0\,.
\eea

We now set up a perturbative expansion in powers of $\Phi$, of the following form:
\begin{multline}
\Phi(r) = \sum_{m=0}^{+\infty}\tilde{V}^{2m+1} \Phi^{(2m+1)}(r)\,,\quad f(r)=f^{(0)}(r)\left[1+\sum_{m=1}^{+\infty}\tilde{V}^{2m}f^{(2m)}(r)\right]\,,\\
a_t(r)=a^{(0)}_t(r)\left[1+\sum_{m=1}^{+\infty}\tilde{V}^{2m}a_t^{(2m)}(r)\right]\,,\quad \delta(r)=\sum_{m=1}^{+\infty}\tilde{V}^{2m}\delta^{(2m)}(r)\,,
\end{multline}
where $\tilde{V} = V/r_+$.

At linear order in $\tilde{V}$, only $\Phi^{(1)}$ is nontrivial, and is given by:
\begin{equation}
\Phi^{(1)}(r) = \phi\left(\frac{r}{r_+}\right)\,,
\end{equation}
with
\begin{equation}
\phi(z) = \frac{(-1)^{-n}}{\,_2F_1\left(\frac{n+1}{2},\frac{n+1}{2}+\frac{1}{2},n+\frac{3}{2},-\frac{1}{2}\right)}\frac{(1-z)^n}{(2+z)^{n+1}}\, _2F_1\left(\frac{n+1}{2},\frac{n+1}{2}+\frac{1}{2},n+\frac{3}{2},-\frac{(1-z)^2}{2 (z+2)^2}\right)\,.
\end{equation}
where we have defined (note $n$ is not necessarily integer)
\be
k^2/r_+^2 \equiv 2(1+3n+3n^2) \,.
\ee
One can readily integrate for the second order metric and gauge field perturbations $f^{(2)}(r)$, $\delta^{(2)}(r)$ and $a_t^{(2)}(r)$, but their expressions are too lengthy to be presented here.

One can now proceed perturbatively and compute the corrections to $\eta/s$ in powers of $\tilde{V}$. The first nontrivial correction to $h_o$ will appear at second order in $\tilde{V}$. Namely, we set
\begin{equation}
h_o = \sum_{m=0}^{+\infty} h_o^{(2m)}\left(\frac{r}{r_+}\right) \tilde{V}^{2m}\,.
\end{equation}
At zeroth order one finds the usual result $h^{(0)}_o(z)=1$, while at second order $h^{(2)}(z)$ can be expressed as a double integral of $\phi(z)^2$, where we have imposed as boundary condition $\lim_{z\to+\infty} h^{(2)}(z)=0$ and regularity at the horizon
\be
h^{(2)}(z) = -8(1+3n+3n^2)\int_z^{+\infty}\mathrm{d}y\frac{1}{3-4y+y^4}\int_1^y \mathrm{d}x\,\phi(x)^2\,.
\ee
Note that the integrals converge.
From (\ref{eq:h2}) we can readily compute $\eta/s$, and we find:
\begin{equation}
4\pi \frac{\eta}{s} =1+2h^{(2)}(1) \tilde{V}^2+\mathcal{O}(\tilde{V}^2)\,.
\end{equation}
We have not found a way to evaluate $h^{(2)}(1)$ for general $n$, but for a given $n$ the integral can be evaluated exactly. For instance, for $n=1$ one finds
\begin{multline}
h^{(2)}(1)_{n=1} = \frac{7}{972 \left(2-2\sqrt{2} \lambda+\lambda^2\right)} \Bigg\{-64 \sqrt{6} \, _4F_3\left(\{\frac{1}{2},\frac{1}{2},\frac{1}{2},\frac{1}{2}\},\{\frac{3}{2},\frac{3}{2},\frac{3}{2}\},\frac{1}{3}\right)-48  \text{Cl}_2\left(\lambda\right)\\
+80 \sqrt{2} \text{Cl}_2\left(\lambda\right)-72\text{Cl}_3\left(\lambda\right)+72 \zeta (3)+88+6 \sqrt{2} \pi  \log ^2\left(\frac{4}{3}\right)+32 \sqrt{2} \arcsin\left(\frac{1}{\sqrt{3}}\right)^3+\\
24 \sqrt{2} \arctan\left(\sqrt{2}\right) \lambda^2-55 \lambda^2+24\sqrt{2} \arctan\left(\sqrt{2}\right)^2 \lambda+12 \sqrt{2} \lambda+16 \sqrt{2} \arctan\left(\sqrt{2}\right)^3\\
-24 \sqrt{2} \pi  \arctan\left(\sqrt{2}\right)^2-8 \sqrt{2} \pi ^2 \mathrm{arccot}\left(\frac{5}{\sqrt{2}}\right)-6 \sqrt{2} \log^2\left(\frac{4}{3}\right) \lambda-12 \sqrt{2} \log ^2\left(\frac{4}{3}\right) \arctan\left(\sqrt{2}\right)\\
-6 \log\left(\frac{4}{3}\right) \lambda^2+40 \sqrt{2} \log \left(\frac{4}{3}\right) \lambda\Bigg\}\,,
\end{multline}
where $\lambda = \arctan(2\sqrt{2})$, $\, _4F_3(\{a_1,a_2,a_3,a_4\},\{b_1,b_2,b_3\},z)$ is a generalized hypergeometric function and the $\text{Cl}_n(\theta)$ and $\text{Sl}_n(\theta)$ are the standard Clausen functions.

At large $k$, or equivalently large $n$, the integral can also be evaluated using a saddle point approximation. The result turns out to take a remarkably simple form:
\begin{equation}
4\pi\frac{\eta}{s} = 1-\frac{V^2}{k^2}+\mathcal{O}\left[\left(\frac{V}{k}\right)^3\right]\,.
\end{equation}

\section{\texorpdfstring{$\eta/s$}{test} in massive gravity}
\label{sec:massive}

The simplest theory of `massive gravity' that has been used to study momentum relaxation
in holography is \cite{Vegh:2013sk}
  \begin{align}
    S ={}& \frac{1}{16 \pi G_N} \int \dd^4 x \, \sqrt{{-} g} \left( R +
      \frac{6}{L^2} + \alpha \, \tr {\cal K} \right)\, ,
  \end{align}
  where ${\cal K}^a{}_b = \sqrt{g^{ac} f_{cb}}$ and $f_{\mu \nu}$ is
  the reference metric which is taken to be $f=\diag(0,0,1,1)$.
 We are restricting to the neutral theory, so there is no Maxwell field. 
  The equations of motion following from this action are (setting $L=1$):
  \begin{align}
    R_{ab} - \frac{1}{2} R g_{ab} - 3 g_{ab} ={}& {-}
    \frac{1}{2} \alpha \left( {\cal K}_{ab} - \tr {\cal K} \, g_{ab}
    \right)\, .\label{eq:masss}
  \end{align}
 These admit the simple black brane solution
   \begin{align}
    \dd s^2 ={}& \frac{1}{r^2} \left( {-} f(r) \dd t^2 + \frac{\dd
        r^2}{f(r)} + \dd x^2 + \dd y^2 \right)\, ,
  \end{align}
  where the emblackening factor reads:
  \begin{align}
    f(r) ={}& 1 - \frac{r^3}{r_+^3} - \frac{\alpha}{2} r \left( 1 -
      \frac{r^2}{r_+^2} \right)\, .
  \end{align}
  The horizon is located at $r=r_+$ and has temperature
  \begin{align}
    4 \pi T ={}& \frac{3+ r_+ \alpha}{r_+}\, .
  \end{align}
  At $T=0$, the horizon degenerates with $\alpha = \frac{-3}{r_+}$,
  yielding a near-horizon $AdS_2 \times \R^2$ region, where the emblackening
  factor takes the form:
  \begin{align}
    f_{T=0}(r) = \frac{1}{2} \left(\frac{r}{r_+} - 1\right)^2
    \left(\frac{r}{r_+} + 2 \right)\, .
  \end{align}
  
  To evaluate the shear viscosity, we need to look at the metric
  perturbations about this background. From the equations of motion
  (\ref{eq:masss}), the wave equation for shear metric fluctuations
  $(\delta g)^x{}_y = h_o(r)$ is found to be
  \be
    \Box h_o ={} m^2(r) h_o \,, \qquad \qquad m^2(r)={} \frac{{-} \alpha r}{2}\, . \label{eq:box}
  \ee
  Positive $m^2$ corresponds to $\alpha<0$, which is the
  region of stability discussed in \cite{Vegh:2013sk}. The mass is nonzero
  on the horizon, even at zero temperature.
  
At $T=0$, using the form of $f(r)$ above, the near-horizon wave equation becomes
  \begin{align}
    \frac{1}{\sqrt{{-} g}} \partial_r \left( \sqrt{{-} g}
      g^{rr} \partial_r h_o \right) = \frac{3}{2} \partial_\rho
    \left( \rho^2 \partial_\rho h_o \right) = \frac{3}{2} h_o \, ,
  \end{align}
  where we have set $\rho=\frac{r}{r_+} - 1$. Assuming $h_o \sim
  \rho^\nu$ as $\rho \to 0$, one easily finds:
  \begin{align}
    2 \nu ={}& \sqrt{5} - 1 \simeq 1.236\, .
  \end{align}
  Therefore, as discussed in the main text, we expect that at low temperatures
  \be
  \frac{4\pi \eta}{s} \sim T^{\sqrt{5}-1} \,.
  \ee
  We have verified this expectation by solving the full equation (\ref{eq:box}) numerically.


\begin{thebibliography}{99}

\bibitem{Kovtun:2004de} 
  P.~Kovtun, D.~T.~Son and A.~O.~Starinets,
  ``Viscosity in strongly interacting quantum field theories from black hole physics,''
  Phys.\ Rev.\ Lett.\  {\bf 94}, 111601 (2005)
  [hep-th/0405231].
  
\bibitem{Policastro:2001yc} 
  G.~Policastro, D.~T.~Son and A.~O.~Starinets,
  ``The Shear viscosity of strongly coupled N=4 supersymmetric Yang-Mills plasma,''
  Phys.\ Rev.\ Lett.\  {\bf 87}, 081601 (2001)
  [hep-th/0104066].
  
\bibitem{Kovtun:2003wp} 
  P.~Kovtun, D.~T.~Son and A.~O.~Starinets,
  ``Holography and hydrodynamics: Diffusion on stretched horizons,''
  JHEP {\bf 0310}, 064 (2003)
  [hep-th/0309213].
  
\bibitem{Adams:2012th} 
  A.~Adams, L.~D.~Carr, T.~SchŠfer, P.~Steinberg and J.~E.~Thomas,
  ``Strongly Correlated Quantum Fluids: Ultracold Quantum Gases, Quantum Chromodynamic Plasmas, and Holographic Duality,''
  New J.\ Phys.\  {\bf 14}, 115009 (2012)
  [arXiv:1205.5180 [hep-th]].
  
\bibitem{Hartnoll:2014lpa} 
  S.~A.~Hartnoll,
  ``Theory of universal incoherent metallic transport,''
  Nature Phys.\  {\bf 11}, 54 (2015)
  doi:10.1038/nphys3174
  [arXiv:1405.3651 [cond-mat.str-el]].
  
\bibitem{Lucas:2015lna} 
  A.~Lucas,
  ``Hydrodynamic transport in strongly coupled disordered quantum field theories,''
  New J.\ Phys.\  {\bf 17}, no. 11, 113007 (2015)
  [arXiv:1506.02662 [hep-th]].
  
\bibitem{Grozdanov:2015qia} 
  S.~Grozdanov, A.~Lucas, S.~Sachdev and K.~Schalm,
  ``Absence of disorder-driven metal-insulator transitions in simple holographic models,''
  Phys.\ Rev.\ Lett.\  {\bf 115}, no. 22, 221601 (2015)
  [arXiv:1507.00003 [hep-th]].
  
\bibitem{Sekino:2008he} 
  Y.~Sekino and L.~Susskind,
  ``Fast Scramblers,''
  JHEP {\bf 0810}, 065 (2008)
  [arXiv:0808.2096 [hep-th]].
  
\bibitem{Maldacena:2015waa} 
  J.~Maldacena, S.~H.~Shenker and D.~Stanford,
  ``A bound on chaos,''
  arXiv:1503.01409 [hep-th].
  
\bibitem{Kats:2007mq} 
  Y.~Kats and P.~Petrov,
  ``Effect of curvature squared corrections in AdS on the viscosity of the dual gauge theory,''
  JHEP {\bf 0901}, 044 (2009)
  doi:10.1088/1126-6708/2009/01/044
  [arXiv:0712.0743 [hep-th]].
  
\bibitem{Buchel:2008vz} 
  A.~Buchel, R.~C.~Myers and A.~Sinha,
  ``Beyond eta/s = 1/4 pi,''
  JHEP {\bf 0903}, 084 (2009)
  [arXiv:0812.2521 [hep-th]].
  
\bibitem{Brigante:2008gz} 
  M.~Brigante, H.~Liu, R.~C.~Myers, S.~Shenker and S.~Yaida,
  ``The Viscosity Bound and Causality Violation,''
  Phys.\ Rev.\ Lett.\  {\bf 100}, 191601 (2008)
  [arXiv:0802.3318 [hep-th]].
  
\bibitem{Hofman:2008ar} 
  D.~M.~Hofman and J.~Maldacena,
  ``Conformal collider physics: Energy and charge correlations,''
  JHEP {\bf 0805}, 012 (2008)
  [arXiv:0803.1467 [hep-th]].
  
\bibitem{Cremonini:2011iq} 
  S.~Cremonini,
  ``The Shear Viscosity to Entropy Ratio: A Status Report,''
  Mod.\ Phys.\ Lett.\ B {\bf 25}, 1867 (2011)
  [arXiv:1108.0677 [hep-th]].
  
\bibitem{Rebhan:2011vd} 
  A.~Rebhan and D.~Steineder,
  ``Violation of the Holographic Viscosity Bound in a Strongly Coupled Anisotropic Plasma,''
  Phys.\ Rev.\ Lett.\  {\bf 108}, 021601 (2012)
  [arXiv:1110.6825 [hep-th]].
  
\bibitem{Mamo:2012sy} 
  K.~A.~Mamo,
  ``Holographic RG flow of the shear viscosity to entropy density ratio in strongly coupled anisotropic plasma,''
  JHEP {\bf 1210}, 070 (2012)
  [arXiv:1205.1797 [hep-th]].
  
\bibitem{Jain:2014vka} 
  S.~Jain, N.~Kundu, K.~Sen, A.~Sinha and S.~P.~Trivedi,
  ``A Strongly Coupled Anisotropic Fluid From Dilaton Driven Holography,''
  JHEP {\bf 1501}, 005 (2015)
  [arXiv:1406.4874 [hep-th]].
  
\bibitem{Critelli:2014kra} 
  R.~Critelli, S.~I.~Finazzo, M.~Zaniboni and J.~Noronha,
  ``Anisotropic shear viscosity of a strongly coupled non-Abelian plasma from magnetic branes,''
  Phys.\ Rev.\ D {\bf 90}, no. 6, 066006 (2014)
  [arXiv:1406.6019 [hep-th]].
  
\bibitem{Jain:2015txa} 
  S.~Jain, R.~Samanta and S.~P.~Trivedi,
  ``The Shear Viscosity in Anisotropic Phases,''
  JHEP {\bf 1510}, 028 (2015)
  [arXiv:1506.01899 [hep-th]].
  
 \bibitem{anis}
 M.~Koschorreck, D.~Pertot, E.~Vogt, and M.~K\"ohl, ``Universal spin dynamics in two-dimensional Fermi gases,'' Nat Phys {\bf 9}, 405 (2013).
  
\bibitem{Son:2007vk} 
  D.~T.~Son and A.~O.~Starinets,
  ``Viscosity, Black Holes, and Quantum Field Theory,''
  Ann.\ Rev.\ Nucl.\ Part.\ Sci.\  {\bf 57}, 95 (2007)
  [arXiv:0704.0240 [hep-th]].
  
\bibitem{Iqbal:2008by} 
  N.~Iqbal and H.~Liu,
  ``Universality of the hydrodynamic limit in AdS/CFT and the membrane paradigm,''
  Phys.\ Rev.\ D {\bf 79}, 025023 (2009)
  [arXiv:0809.3808 [hep-th]].

\bibitem{zaanen}
J.~Zaanen, ``Superconductivity: Why the temperature is high,'' Nature, {\bf 430}, 512 (2004).

\bibitem{Donos:2014cya} 
  A.~Donos and J.~P.~Gauntlett,
  ``Thermoelectric DC conductivities from black hole horizons,''
  JHEP {\bf 1411}, 081 (2014)
  [arXiv:1406.4742 [hep-th]].
    
\bibitem{Hartnoll:2009sz} 
  S.~A.~Hartnoll,
  ``Lectures on holographic methods for condensed matter physics,''
  Class.\ Quant.\ Grav.\  {\bf 26}, 224002 (2009)
  [arXiv:0903.3246 [hep-th]].
  
\bibitem{Hartnoll:2014cua}
  S.~A.~Hartnoll and J.~E.~Santos,
  ``Disordered horizons: Holography of randomly disordered fixed points,''
  Phys.\ Rev.\ Lett.\  {\bf 112} (2014) 231601
  [arXiv:1402.0872 [hep-th]].

\bibitem{Hartnoll:2015faa}
  S.~A.~Hartnoll, D.~M.~Ramirez and J.~E.~Santos,
  ``Emergent scale invariance of disordered horizons,''
  JHEP {\bf 1509} (2015) 160
  [arXiv:1504.03324 [hep-th]].

\bibitem{Hartnoll:2015rza}
  S.~A.~Hartnoll, D.~M.~Ramirez and J.~E.~Santos,
  ``Thermal conductivity at a disordered quantum critical point,''
  arXiv:1508.04435 [hep-th].

\bibitem{Hartnoll:2014gaa}
  S.~A.~Hartnoll and J.~E.~Santos,
  ``Cold planar horizons are floppy,''
  Phys.\ Rev.\ D {\bf 89} (2014) 12,  126002
  [arXiv:1403.4612 [hep-th]].

\bibitem{Donos:2012js} 
  A.~Donos and S.~A.~Hartnoll,
  ``Interaction-driven localization in holography,''
  Nature Phys.\  {\bf 9}, 649 (2013)
  [arXiv:1212.2998].
  
\bibitem{Donos:2013eha} 
  A.~Donos and J.~P.~Gauntlett,
  ``Holographic Q-lattices,''
  JHEP {\bf 1404}, 040 (2014)
  [arXiv:1311.3292 [hep-th]].

\bibitem{Andrade:2013gsa} 
  T.~Andrade and B.~Withers,
  ``A simple holographic model of momentum relaxation,''
  JHEP {\bf 1405}, 101 (2014)
  [arXiv:1311.5157 [hep-th]].
  
\bibitem{analytis}  
I.~M.~Hayes, N.~P.~Breznay, T.~Helm, P.~Moll, M.~Wartenbe, R.~D.~McDonald, A.~Shekhter and J.~G.~Analytis, ``Magnetoresistance near a quantum critical point,'' arXiv:1412.6484 [cond-mat].

\bibitem{Donos:2015gia} 
  A.~Donos and J.~P.~Gauntlett,
  ``Navier-Stokes Equations on Black Hole Horizons and DC Thermoelectric Conductivity,''
  arXiv:1506.01360 [hep-th].
  
\bibitem{Lucas:2015vna} 
  A.~Lucas,
  ``Conductivity of a strange metal: from holography to memory functions,''
  JHEP {\bf 1503}, 071 (2015)
  [arXiv:1501.05656 [hep-th]].
  
\bibitem{Chakrabarti:2010xy}
  S.~K.~Chakrabarti, S.~Chakrabortty and S.~Jain,
  ``Proof of universality of electrical conductivity at finite chemical potential,''
  JHEP {\bf 1102} (2011) 073
  [arXiv:1011.3499 [hep-th]].
  
\bibitem{Davison:2015taa} 
  R.~A.~Davison, B.~GoutŽraux and S.~A.~Hartnoll,
  ``Incoherent transport in clean quantum critical metals,''
  JHEP {\bf 1510}, 112 (2015)
  [arXiv:1507.07137 [hep-th]].
  
\bibitem{Vegh:2013sk} 
  D.~Vegh,
  ``Holography without translational symmetry,''
  arXiv:1301.0537 [hep-th].
  
\bibitem{Blake:2013owa} 
  M.~Blake, D.~Tong and D.~Vegh,
  ``Holographic Lattices Give the Graviton an Effective Mass,''
  Phys.\ Rev.\ Lett.\  {\bf 112}, no. 7, 071602 (2014)
  [arXiv:1310.3832 [hep-th]].
  
\bibitem{Edalati:2009bi} 
  M.~Edalati, J.~I.~Jottar and R.~G.~Leigh,
  ``Transport Coefficients at Zero Temperature from Extremal Black Holes,''
  JHEP {\bf 1001}, 018 (2010)
  [arXiv:0910.0645 [hep-th]].
  
   \bibitem{Hawking:1974rv} 
  S.~W.~Hawking,
  ``Black hole explosions,''
  Nature {\bf 248}, 30 (1974).
  
    \bibitem{deHaro:2000vlm} 
  S.~de Haro, S.~N.~Solodukhin and K.~Skenderis,
  ``Holographic reconstruction of space-time and renormalization in the AdS / CFT correspondence,''
  Commun.\ Math.\ Phys.\  {\bf 217}, 595 (2001)
  [hep-th/0002230].
  
  \bibitem{Horowitz:1998ha} 
  G.~T.~Horowitz and R.~C.~Myers,
  ``The AdS / CFT correspondence and a new positive energy conjecture for general relativity,''
  Phys.\ Rev.\ D {\bf 59}, 026005 (1998)
  [hep-th/9808079].


\bibitem{Davison:2014lua} 
  R.~A.~Davison and B.~GoutŽraux,
  ``Momentum dissipation and effective theories of coherent and incoherent transport,''
  JHEP {\bf 1501}, 039 (2015)
  [arXiv:1411.1062 [hep-th]].

\bibitem{Gubser:2008wz} 
  S.~S.~Gubser and F.~D.~Rocha,
  ``The gravity dual to a quantum critical point with spontaneous symmetry breaking,''
  Phys.\ Rev.\ Lett.\  {\bf 102}, 061601 (2009)
  [arXiv:0807.1737 [hep-th]].

\bibitem{Faulkner:2011tm} 
  T.~Faulkner, N.~Iqbal, H.~Liu, J.~McGreevy and D.~Vegh,
  ``Holographic non-Fermi liquid fixed points,''
  Phil.\  Trans.\  Roy.\  Soc.\ A {\bf  369}, 1640 (2011)
  [arXiv:1101.0597 [hep-th]].

\bibitem{Donos:2012ra} 
  A.~Donos and S.~A.~Hartnoll,
  ``Universal linear in temperature resistivity from black hole superradiance,''
  Phys.\ Rev.\ D {\bf 86}, 124046 (2012)
  [arXiv:1208.4102 [hep-th]].
  


 
  

  
  


\bibitem{Kachru:2008yh} 
  S.~Kachru, X.~Liu and M.~Mulligan,
  ``Gravity duals of Lifshitz-like fixed points,''
  Phys.\ Rev.\ D {\bf 78}, 106005 (2008)
  [arXiv:0808.1725 [hep-th]].
  
\bibitem{Hartnoll:2011fn} 
  S.~A.~Hartnoll,
  ``Horizons, holography and condensed matter,''
  arXiv:1106.4324 [hep-th].

\bibitem{Headrick:2009pv} 
  M.~Headrick, S.~Kitchen and T.~Wiseman,
  ``A New approach to static numerical relativity, and its application to Kaluza-Klein black holes,''
  Class.\ Quant.\ Grav.\  {\bf 27}, 035002 (2010)
  [arXiv:0905.1822 [gr-qc]].
  
  \bibitem{Figueras:2011va} 
  P.~Figueras, J.~Lucietti and T.~Wiseman,
  ``Ricci solitons, Ricci flow, and strongly coupled CFT in the Schwarzschild Unruh or Boulware vacua,''
  Class.\ Quant.\ Grav.\  {\bf 28}, 215018 (2011)
  [arXiv:1104.4489 [hep-th]].
  
  \bibitem{Dias:2015nua} 
  O.~J.~C.~Dias, J.~E.~Santos and B.~Way,
  ``Numerical Methods for Finding Stationary Gravitational Solutions,''
  arXiv:1510.02804 [hep-th].

\bibitem{Son:2002sd} 
  D.~T.~Son and A.~O.~Starinets,
  ``Minkowski space correlators in AdS / CFT correspondence: Recipe and applications,''
  JHEP {\bf 0209}, 042 (2002)
  [hep-th/0205051].
  
\bibitem{Hartnoll:2012rj} 
  S.~A.~Hartnoll and D.~M.~Hofman,
  ``Locally Critical Resistivities from Umklapp Scattering,''
  Phys.\ Rev.\ Lett.\  {\bf 108}, 241601 (2012)
  [arXiv:1201.3917 [hep-th]].

\bibitem{Horowitz:2009ij} 
  G.~T.~Horowitz and M.~M.~Roberts,
  ``Zero Temperature Limit of Holographic Superconductors,''
  JHEP {\bf 0911}, 015 (2009)
  [arXiv:0908.3677 [hep-th]].
 
  
  \bibitem{ziman}
 J.~M.~Ziman, ``The General Variational Principle of Transport Theory,'' Can.~J.~Phys. {\bf 34}, 1256 (1956).

\bibitem{bounds}
K.~Van Acoleyen, M.~Mari\"en, and F.~Verstraete, ``Entanglement Rates and Area Laws,''
Phys. Rev. Lett. {\bf 111}, 170501 (2013).

\bibitem{Hartman:2015apr} 
  T.~Hartman and N.~Afkhami-Jeddi,
  ``Speed Limits for Entanglement,''
  arXiv:1512.02695 [hep-th].
  
\bibitem{Casini:2012ei} 
  H.~Casini and M.~Huerta,
  ``On the RG running of the entanglement entropy of a circle,''
  Phys.\ Rev.\ D {\bf 85}, 125016 (2012)
  [arXiv:1202.5650 [hep-th]].

\bibitem{Casini:2015woa} 
  H.~Casini, M.~Huerta, R.~C.~Myers and A.~Yale,
  ``Mutual information and the F-theorem,''
  JHEP {\bf 1510}, 003 (2015)
  [arXiv:1506.06195 [hep-th]].
  
\bibitem{Rattazzi:2008pe} 
  R.~Rattazzi, V.~S.~Rychkov, E.~Tonni and A.~Vichi,
  ``Bounding scalar operator dimensions in 4D CFT,''
  JHEP {\bf 0812}, 031 (2008)
  [arXiv:0807.0004 [hep-th]].

\bibitem{Alberte:2016xja} 
  L.~Alberte, M.~Baggioli and O.~Pujolas,
  ``Viscosity bound violation in holographic solids and the viscoelastic response,''
  arXiv:1601.03384 [hep-th].
  
\bibitem{Burikham:2016roo} 
  P.~Burikham and N.~Poovuttikul,
  ``Shear viscosity in holography and effective theory of transport without translational symmetry,''
  arXiv:1601.04624 [hep-th].

\end{thebibliography}
\end{document}